\documentclass[numberedappendix]{emulateapj}
\usepackage[varg]{txfonts}
\usepackage{paralist}
\usepackage[latin1]{inputenc}

\let\url\relax
\usepackage[hypertex]{hyperref}
\usepackage[all]{hypcap}

\providecommand\toprule{\hline}
\providecommand\midrule{\hline}
\providecommand\bottomrule{\hline}
\providecommand\cmidrule{\cline}

\makeatletter
\renewcommand{\fps@figure}{t!}
\makeatother

\newcounter{ionstage}
\renewcommand{\ion}[2]{\setcounter{ionstage}{#2}%
  \ensuremath{\mathrm{#1\,\scriptstyle\Roman{ionstage}}}}
\newcommand\Sub[1]{_\mathrm{#1}}

\newcommand\unit[2]{\ensuremath{#1\ \mathrm{#2}}}
\newcommand\reciprocal[1]{\ensuremath{#1^{-1}}}
\newcommand\rpsquare[1]{\ensuremath{#1^{-2}}}
\newcommand\rpcubic[1]{\ensuremath{#1^{-3}}}
\renewcommand\square[1]{\ensuremath{#1^{2}}}
\newcommand\centi{c}
\newcommand\milli{m}
\newcommand\kilo{k}
\newcommand\second{s}
\newcommand\meter{m}
\newcommand\angstrom{\AA}
\newcommand\kelvin{K}
\newcommand\usk{\ }
\newcommand{\Density}[1]{\unit{#1}{\rpcubic{\centi\meter}}}
\newcommand{\Velocity}[1]{\unit{#1}{\kilo\meter\usk\reciprocal\second}}

\newcommand{\Wav}[1]{\unit{\lambda#1}{\textrm{\angstrom}}}
\newcommand\erg{\mathrm{erg}}
\newcommand\Emissivity[1]{\unit{#1}{\erg\usk\rpcubic{\centi\meter}\usk\reciprocal\second}}

\newcommand\hii{\ion{H}{2}}
\newcommand{\OI}{[\ion{O}{1}]}
\newcommand{\OIlam}{\OI{}\,\Wav{6300}}
\newcommand{\NII}{[\ion{N}{2}]}
\newcommand{\NIIlam}{\NII{}\,\Wav{6584}}

\newcommand{\OIII}{[\ion{O}{3}]}
\newcommand{\OIIIlam}{\ensuremath{\OIII{}\,\Wav{5007}}}

\newcommand{\SII}{[\ion{S}{2}]}
\newcommand{\SIIlam}{\SII{}\,\Wav{6731}}
\newcommand{\SIIbothlam}{\SII{}\,\Wav{6716+6731}}
\newcommand{\Ha}{\ensuremath{\mathrm{H}\alpha}}
\newcommand{\Halam}{\ensuremath{\Ha{}\,\Wav{6563}}}
\newcommand{\Hb}{\ensuremath{\mathrm{H}\beta}}

\newlength\EditNoteWidth
\newcommand{\thC}{$\theta^1\,$Ori\,C\@}

\newcommand{\lamad}{\ensuremath{\lambda\Sub{ad}}}
\newcommand{\xiad}{\ensuremath{\xi\Sub{ad}}}
\newcommand{\Mach}{\ensuremath{\mathcal{M}}}
\newcommand{\Ne}{\ensuremath{n_\mathrm{e}}}

\newcommand{\Col}{_\mathrm{col}}
\newcommand{\Rec}{_\mathrm{rec}}
\newcommand{\Max}{_\mathrm{m}} 

\newcommand{\subH}{_\mathrm{\scriptscriptstyle H}}

\newcommand{\sigbar}{\bar{\sigma}}

\begin{document}

\title{Self-Consistent Dynamic Models of Steady Ionization Fronts:\\
  I.~Weak-D and Weak-R Fronts}

\author{W. J. Henney\altaffilmark{1} and S. J. Arthur\altaffilmark{1}} %
\affil{Centro de Radioastronomía y Astrofísica, UNAM Campus Morelia,
  Apdo.\@ Postal 3-72, 58090 Morelia, Michoacán, México.}
\email{w.henney,j.arthur@astrosmo.unam.mx} 
\author{R. J. R. Williams} %
\affil{Atomic Weapons Establishment,  Aldermaston, RG7~4PR, UK.} %
\email{Robin.Williams@awe.co.uk}
\author{G. J. Ferland} %
\affil{Department of Physics \& Astronomy, University of Kentucky, 177
  Chem.-Phys.\@ Building, 600 Rose Street Lexington, Kentucky
  40506-0055, USA.} %
\email{gary@pop.uky.edu}
\altaffiltext{1}{Work
  carried out in part while on sabbatical at Department of Physics and
  Astronomy, University of Leeds, LS2~9JT, UK.}

\begin{abstract}
  We present a method for including steady-state gas flows in the
  plasma physics code Cloudy, which was previously restricted to
  modeling static configurations. The numerical algorithms are
  described in detail, together with an example application to
  plane-parallel ionization-bounded \hii{} regions. As well as
  providing the foundation for future applications to more complex
  flows, we find the following specific results regarding the effect
  of advection upon ionization fronts in \hii{} regions:

  1. Significant direct effects of advection on the global emission
    properties occur only when the ionization parameter is lower than
    is typical for \hii{} regions. For higher ionization parameters,
    advective effects are indirect and largely confined to the
    immediate vicinity of the ionization front.

  2 The overheating of partially ionized gas in the front is not
    large, even for supersonic (R-type) fronts. For subsonic (D-type)
    fronts we do not find the temperature spike that has been
    previously claimed. 

  3. The most significant morphological signature of advective
    fronts is an electron density spike that occurs at the ionization
    front whenever the relative velocity between the ionized gas and
    the front exceeds about one half the ionized isothermal sound
    speed.  Observational evidence for such a spike is found in
    \NIIlam{} images of the Orion bar.

  4. Plane-parallel, weak-D fronts are found to show at best a
    shallow correlation between mean velocity and ionization potential
    for optical emission lines even when the flow velocity closely
    approaches the ionized sound speed. Steep gradients in velocity
    versus ionization, such as those observed in the Orion nebula,
    seem to require transonic flows, which are only possible in a
    diverging geometry when radiation forces are included.
\end{abstract}


\shorttitle{Dynamic Models of Ionization Fronts}

\section{Introduction}
\label{sec:introduction}

The classic early work on the effects of dynamics on the emission
structure of \hii{} regions was carried out by
\citet{1977MNRAS.179...63H}, who studied weak-D fronts in which the
gas motions are always subsonic with respect to the front. Within the
fully ionized interior of the \hii{} region the gas was found to be
close to thermal and ionization equilibrium.  Significant
non-equilibrium effects induced by the dynamics are confined to the
edge of the region, near the ionization front, where there exist large
gradients in the radiation-field intensity and in the physical
conditions of the gas such as temperature and degree of ionization.
\citeauthor{1977MNRAS.179...63H} found that the dynamics only had a
small effect on the integrated forbidden line spectrum of the models
he considered. However, there are various reasons to revisit such
calculations now.

First, ionization fronts are now studied in a diverse range of
astrophysical contexts in which the classical, spherically symmetric,
subsonic expansion studied by \citeauthor{1977MNRAS.179...63H} may be
the exception rather than the rule. Transonic photoevaporation flows
seem to be a ubiquitous feature of photoionized regions, ranging in
scale from cometary knots in planetary nebula
\citep{2001ApJ...548..288L} and photoevaporated circumstellar disks in
\hii{} regions \citep[ and references therein]{2001ARA&A..39...99O} up
to champagne flows in giant extragalactic \hii{} regions
\citep{1998AJ....116..163S} and the photoevaporation of cosmological
mini halos \citep{2001RMxAC..10..109S}. In such flows, non-equilibrium
effects will be somewhat more important than in subsonic weak-D fronts
due to the higher velocities involved. \citet{2000ApJ...535..847S}
made a first attempt at detailed modeling of the emission structure
of the flow from the head of the columns in M16, using static
equilibrium models for the ionization structure.

Second, 
continuous improvement in the spatial resolution and wavelength
coverage of observations, together with advances in theoretical and
observational atomic physics, now allow a much more detailed
comparison between model predictions and spatially resolved
observations of a multitude of emitting species. In this context, even
moderate and localized changes in the predicted spectrum due to
dynamical effects can be important.

Third, a dynamical treatment allows the self-consistent calculation of
the velocity field, which allows comparison with high-resolution
spectral line profile observations that provide further constraints on
the models. Furthermore, it permits a unified treatment of the entire
flow from cold, molecular gas, through the photon-dominated or
photodissociation region (PDR), and into the ionized region. Previous
studies of non-equilibrium models of PDRs
\citep{1978ApJ...225..405L,1998A&A...337..517N,1998ApJ...495..853S}
have tended to treat the PDR in isolation without considering the
\hii{} region in any detail. \citet{2000ApJ...539..258R} presented
numerical radiation-hydrodynamic simulations of the photoevaporation
of proplyd disks but the physics of the PDR was calculated in a
simplified manner. 

In common with most other photoionization codes, Cloudy
\citep{1998PASP..110..761F,2000RMxAC...9..153F} has traditionally
calculated static equilibrium models in which time-dependent effects
are neglected, such as isochoric (constant density) or isobaric
(constant pressure) configurations. The task of combining
hydrodynamics with detailed simulation of the plasma microphysics can
be approached from one of two angles. One method would be to add the
atomic physics and radiative transfer processes to an existing
time-dependent hydrodynamic code. The other, which is the method
pursued in this work, is to add steady-state dynamics to an existing
plasma physics code.

The current paper is the first in a series of three that will present
detailed results from our program to include a self-consistent
treatment of steady-state advection in a realistic plasma physics
code. This first paper introduces the methodology employed in the
series as a whole and then concentrates on the restricted problem of
``weak'' ionization fronts in a plane-parallel geometry. The second
paper of the series includes the molecular reaction networks necessary
for modeling the neutral/molecular PDR, while the third paper
considers ``strong D'' ionization fronts, where the gas accelerates
through a sonic point, as found in divergent geometries such as the
photoevaporation of globules.

We first discuss the general problem of advection in ionization fronts
(section~\ref{sec:advect-ioniz-fronts}). We then describe the
modifications that have been made to the Cloudy photoionization code
in order to treat steady-state flow
(section~\ref{sec:adding-dynam-cloudy}). Results from a small sample
of representative models are presented in section~\ref{sec:results}
and the application of our results to observations of the Orion nebula
is discussed in section~\ref{sec:discussion}. Further technical
details of the physical processes and computational algorithms are
presented in a series of appendices.

\section{Advection in Ionization Fronts}
\label{sec:advect-ioniz-fronts}

The classification of ionization fronts depends on the behavior of the
gas velocity in the frame of reference in which the ionization front
is fixed \citep{1954BAN....12..187K,Goldsworthy-1961}. In this frame,
the gas flows from the neutral side of the front (denoted
\emph{upstream}) towards the ionized side (denoted \emph{downstream}).
If the upstream gas velocity on the far neutral side is subsonic with
respect to the front then the front is said to be \emph{D-type}, while
if it is supersonic the front is said to be \emph{R-type}. A further
distinction is made between those fronts that contain an internal
sonic point, which are said to be \emph{strong}, and those that do
not, which are said to be \emph{weak}. For example, a weak-R front
will have supersonic velocities throughout the front, whereas in a
strong-D front the gas starts at subsonic velocities on the neutral
side, accelerating through the front to reach a supersonic exhaust
velocity on the ionized side. When the downstream gas velocity is
exactly sonic the front is said to be \emph{critical}. There is also
the possibility of a \emph{recombination front}, in which the sense of
the gas velocity is reversed and the flow is from the ionized side
towards the neutral side, with a similar range of possible structures
\citep{1968ApJ...151.1145N,1996MNRAS.279..987W}. If the gas is
magnetized, then the classification becomes more complicated since
there are now three wave speeds to take into account (Alfv\`en speed
plus fast and slow magnetosonic speeds) instead of just the sound
speed. Thus, one may have a slow-mode D-critical front, a fast-mode
weak-R front, etc
\citep{1998A&A...331.1099R,2000MNRAS.314..315W,2001MNRAS.325..293W}.

These classification schemes were developed for plane-parallel fronts
but will be approximately valid so long as the radius of curvature of
the front greatly exceeds its thickness. This is usually the case
since ionization fronts are in general very thin compared with the
sizes of \hii{} regions unless the ionization parameter is small (see
below). What type of ionization front actually obtains in a given
situation depends on the upstream and downstream boundary conditions
of the front, in particular the upstream gas density and the
downstream gas pressure and ionizing radiation field, together with
the large-scale geometry of the flow, which need not be
plane-parallel.

Since the gas velocity through the front is high for an R-type front,
so is the flux of neutral particles that must be ionized for a given
upstream density, which in turn requires a high ionizing flux at the
downstream boundary. As a result, R-type fronts are usually transient
phenomena accompanying temporal increases in the ionizing flux, such
as the ``turning on'' of an ionizing source. In the most common case,
the front will be propagating rapidly through slowly moving gas.  In
the limit of an extreme weak-R front, the gas velocity in the
ionization front frame and density are constant throughout the flow.

D-type fronts are more common and the limit of an extreme weak-D front
corresponds to a static constant pressure front in ionization
equilibrium. Weak-D fronts require a high downstream pressure and
therefore are likely to be found in cases where the ionization front
envelops the ionizing source. Strong-D and D-critical fronts, on the
other hand, are consistent with the free escape of the downstream gas
and hence apply to divergent photoevaporation flows, for example, from
globules.

Advection of material through the ionization front may be expected to
have various effects on the emission properties of a nebula. In order
to simplify the discussion, we will consider a plane-parallel nebula,
illuminated at one face by a given radiation field and in a frame of
reference in which the ionization front is at rest. This is
illustrated in Figure~\ref{fig:context}\textit{a}. The gas is supposed
to enter the front from the neutral side with velocity $v\Sub{n}$ and
to leave on the ionized side with velocity $v\Sub{i}$. Results from
this simplified model are described in
\S~\ref{sec:direct-indir-affects} but we first discuss the relation
between this model and real \hii{} regions.

\subsection{Physical Context of Advective Fronts }
\label{sec:phys-cont-advect}

In this section, we consider two typical scenarios in which advective
ionization fronts may be encountered and investigate to what extent
they may approximated as steady flows in the frame of reference of the
ionization front. In this discussion we follow
\citet{1992phas.book.....S} in denoting gas velocities in the frame of
reference of the ionizing star by $u$, gas velocities in the frame of
reference of the ionization front by $v$, and pattern speeds of
ionization and shock fronts by $U$. 

\subsubsection{Classical Strömgren Sphere}
\label{sec:class-stromgr-sphere}
\begin{figure}
  \makebox[0.3\linewidth][l]{(\textit{a})}
  \hfill\makebox[0.3\linewidth][l]{(\textit{b})}
  \hfill\makebox[0.3\linewidth][l]{(\textit{c})}
  \includegraphics[width=0.3\linewidth]{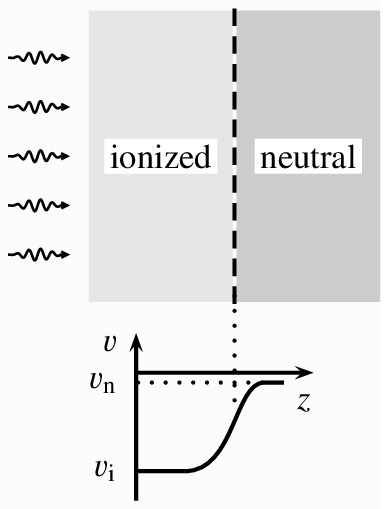}
  \hfill\includegraphics[width=0.3\linewidth]{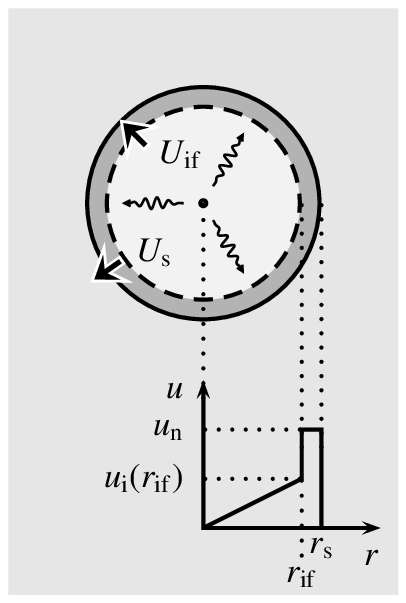}
  \hfill\includegraphics[width=0.3\linewidth]{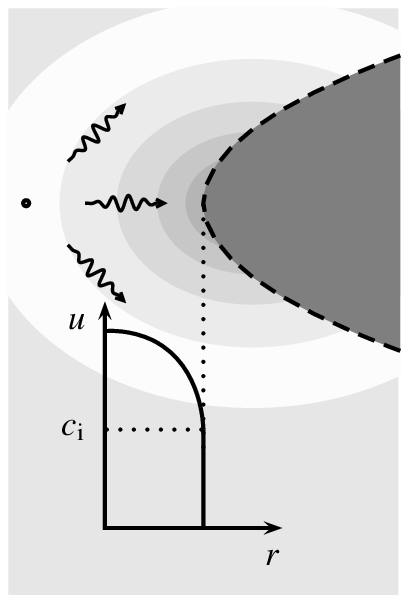}
  \caption{Advective fronts: (\textit{a}) Plane-parallel idealization;
    (\textit{b}) Classical Strömgren sphere; (\textit{c})
    Photoevaporation flow.}
  \label{fig:context}
\end{figure}

The evolution of a classical Strömgren-type \hii{} region in a
constant density medium has been described by many authors
\citep[e.g.,][]{Goldsworthy-1958,1978ppim.book.....S,1992phas.book.....S,1997pism.book.....D}.
If the ionizing source turns on instantaneously, then the ionization
front is initially R-type and propagates supersonically through the
surrounding gas with little accompanying gas motion. By the time the
ionization front reaches the initial Strömgren radius (where the
rate of recombinations in the ionized region approximately balances
the ionizing luminosity of the source), the front propagation velocity
has slowed to the order of the sound speed in the ionized gas and the
front becomes D-type, preceded by a shock that accelerates and
compresses the neutral gas. The initial R-type phase is very short (of
order the recombination timescale), and consequently is of little
observational importance.  The structure of the region during its
subsequent D-type evolution is shown in
Figure~\ref{fig:context}\textit{b}.

\begin{figure}
  \includegraphics[width=\linewidth]{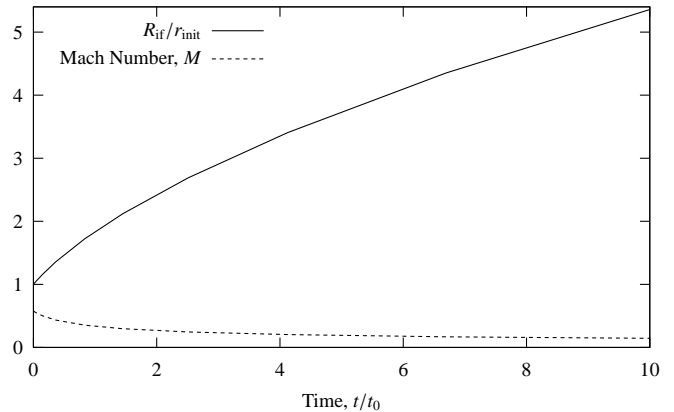}
  \caption{Evolution of the ionization front radius and the
    ``exhaust'' Mach number of newly ionized gas in the
    frame of reference of the advancing ionization front in a
    classical Strömgren sphere. These quantities are plotted as a
    function of time since the front reached the initial Strömgren
    radius, in units of the sound crossing time at that radius.}
  \label{fig:mach-number}
\end{figure}

The propagation speed of the shock is roughly equal to that of the
ionization front and also to the velocity of the gas in the neutral
shell: 
\begin{equation}
  \label{eq:1}
  U\Sub{s} \simeq U\Sub{if} \simeq u\Sub{n} , 
\end{equation}
whereas the ionized gas expands homologously with a velocity that
increases linearly with radius, reaching half the speed of the
ionization front at the front itself: 
\begin{equation}
  \label{eq:2}
  u\Sub{i} = \frac12 \left(\frac{r}{R\Sub{if}}\right) U\Sub{if} . 
\end{equation}
Hence, the velocity of the ionized gas immediately downstream of the
ionization front in the ionization front frame is 
\begin{equation}
  \label{eq:3}
  v\Sub{i} = u\Sub{i}(R\Sub{if}) - U\Sub{if} = -0.5 U\Sub{if} . 
\end{equation}
The ionization front propagates very slightly faster than the gas in
the neutral shell, giving a small upstream neutral gas velocity in the
ionization front frame of
\begin{equation}
  \label{eq:4}
  v\Sub{n} = u\Sub{n} - U\Sub{if} <  - 2 c\Sub{n}^2 / c\Sub{i} \sim 0.1
  \mathrm{\,km\,s}^{-1}, 
\end{equation}
where $c\Sub{n}$, $c\Sub{i}$ are the isothermal sound speeds in the
neutral and ionized gas, respectively.
The evolution of the ionization front propagation speed can be
described in terms of its radius as
\begin{equation}
  \label{eq:5}
  U_\mathrm{if} = \frac{ 2 c\Sub{i} } { \left[
      4\left(R\Sub{if}/R\Sub{init}\right)^{3/2} - 1 \right]^{1/2} } , 
\end{equation} where $R\Sub{init}$ is the initial Strömgren radius.
The Mach number reached by the ionized gas just inside the ionization
front, measured in the frame of reference in which the front is
stationary, is given by $M = 0.5 U\Sub{if} / c\Sub{i}$. This is
plotted in Figure~\ref{fig:mach-number} as a function of ionization
front radius.  During the lifetime of a typical O star (and assuming
an ambient density of order $\Density{1}$), the radius of a classical
\hii{} region will expand by roughly a factor of 4, so, as can be seen
from Figure~\ref{fig:mach-number}, $M = 0.3$--0.5 is typical of the
majority of the evolutionary lifetime.

\subsubsection{Photoevaporation Flow}
\label{sec:phot-flow}

Photoevaporation flows are very common in ionized regions, occurring
whenever the ionization front is convex from the point of view of the
ionizing source, thus allowing the ionized gas to freely stream away
from the front. Examples include bright-rimmed globules and proplyds
in \hii{} regions and cometary knots in planetary nebulae. On a larger
scale, blister-type \hii{} regions can also be considered
photoevaporation flows \citep{1996ApJ...458..222B}.

Figure~\ref{fig:context}\textit{c} show the structure of an idealized
photoevaporation flow, roughly corresponding to the equilibrium
cometary globules of \citet{1990ApJ...354..529B}. The neutral gas is
assumed to be approximately at rest with respect to the ionizing
source and the ionization front to be D-critical with $v\Sub{i}\simeq
u\Sub{i}\simeq c\Sub{i}$. The ionization front eats slowly into the
neutral gas with $U\Sub{i} = - v\Sub{n} = c\Sub{n}/ (2 c\Sub{i}^2 )
\ll c\Sub{i}$, and the Mach number reached by the ionized gas at the
front will be $\simeq 1$.  Outside the front, the ionized gas
accelerates as an approximately isothermal wind, as observed around
the Orion proplyds \citep{1998AJ....116..322H}.

\subsection{Direct and Indirect Effects of Advection}
\label{sec:direct-indir-affects}

In order to provide some physical insight into advective ionization
fronts and to guide the interpretation of the numerical simulations,
we have developed a simple analytic model for a plane-parallel weak-D
ionization front, in which the gas temperature is assumed to be a
prescribed monotonic function of the hydrogen ionization fraction,
$x$. With this assumption, and considering mass and momentum
conservation, it is possible to find algebraic solutions for the
density, gas velocity, and sound speed as functions of $x$ (see
Appendix~\ref{sec:analytic-model}).  These solutions form a
one-parameter family characterized by the maximum Mach number of the
gas in the rest-frame of the ionization front, reached asymptotically
as $x\rightarrow 1$.  If we now take into consideration the ionization
balance and radiative transfer, one can find a solution for $x(z)$,
the ionization fraction as a function of physical depth, by solving a
pair of ordinary differential equations.

\newcommand\IonPar{\ensuremath{\Upsilon}}
Apart from the maximum Mach number, $\Mach\Max$, the solutions are
found to depend on two dimensionless parameters, $\xiad$ and $\tau_*$.
The first of these, $\xiad$ (defined in equation~\ref{eq:xiad}),
is roughly the ratio of recombination length to mean-free-path of
ionizing photons in a D-critical front and is not expected to vary
greatly between \hii{} regions, having a typical value of $\xiad\simeq
10$. The second parameter, $\tau_*$ (defined in
equation~\ref{eq:tau-star}), is roughly the ratio of the thickness
of the fully ionized slab to the thickness of the ionization front and
is proportional to the ionization parameter at the ionized face:
$\tau_* \sim 10^6 \IonPar$, where $\IonPar \equiv F_0 / (n_0 c)$.  The
global importance of advection for the system as a whole can be
characterized by the parameter $\lamad$, defined as the ratio of the
flux of hydrogen atoms through the ionization front to the flux of
ionizing photons at the illuminated face of the slab
(equations~\ref{eq:def-lambda} and
\ref{eq:lamad-xiad-taustar-mach}). For small values of $\lamad$, its
value can be approximated as $\lamad \simeq \xiad \Mach\Max / \tau_*$.
This can be understood as follows: the local effects of advection at
the ionization front itself are always substantial (so long as
$\Mach\Max$ is not too small), being of order $\xiad \Mach\Max$. On
the other hand, the partially-ionized zone occupies only a small
volume compared with the fully ionized gas unless the ionization
parameter is small, so the global effects of advection are reduced by
a factor of $\tau_*$.

\begin{figure}\centering
  \makebox[\linewidth][l]{(\textit{a})}\\[\smallskipamount]
  \includegraphics[width=\linewidth]{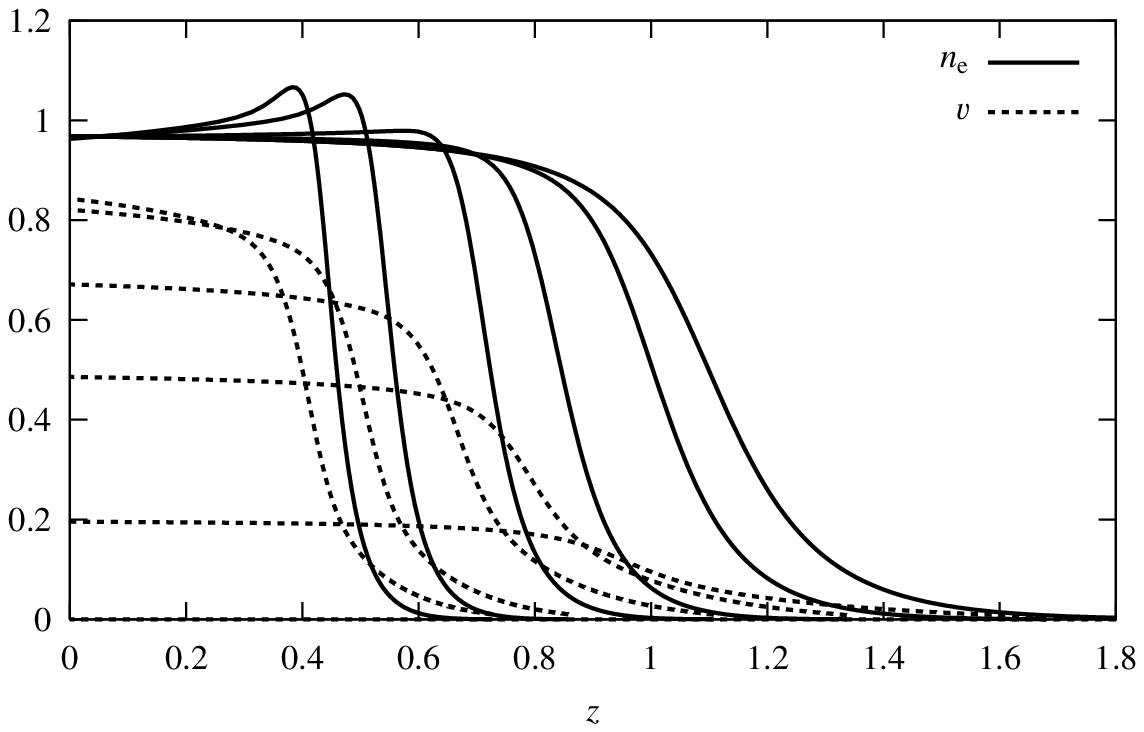}\\
  \makebox[\linewidth][l]{(\textit{b})}\\[\smallskipamount]
  \includegraphics[width=\linewidth]{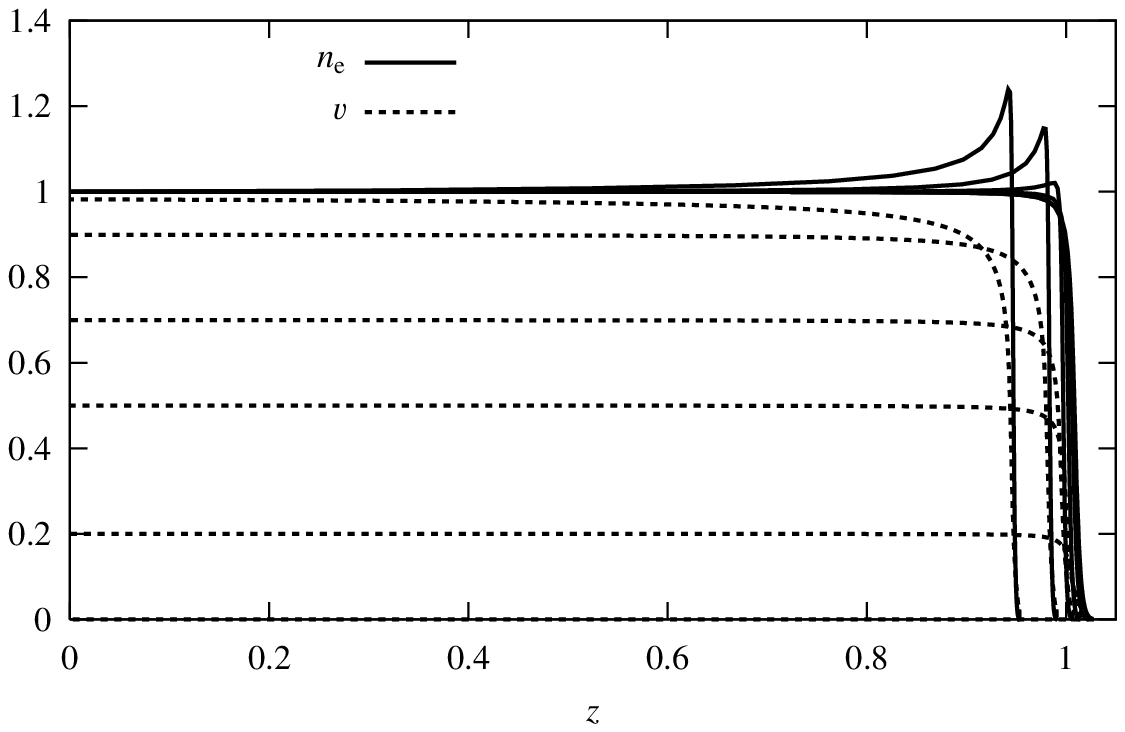}
  \caption{Dynamic slab solutions from the simplified model developed
    in Appendix~\ref{sec:analytic-model}.  (\textit{a}) $\tau_* = 30$,
    low ionization parameter, fat IF\@. (\textit{b}) $\tau_* = 3000$,
    high ionization parameter, thin IF\@. Electron density (solid
    line) and gas velocity (dashed line) are shown in each case for 6
    models with $\Mach\Max = 0.0$, 0.2, 0.5, 0.7, 0.9, 0.99
    (corresponding respectively to successive curves from right to
    left). Electron density is normalized by the fiducial density
    $n\Max$,  velocity is normalized by the maximum sound speed
    $c\Max$, and distance is normalized by the static Strömgren
    depth $z_0$---see Appendix~\ref{sec:analytic-model} for details.}
  \label{fig:wdz}
\end{figure}
Detailed results are calculated in Appendix~\ref{sec:analytic-model}
for both a low ionization parameter model ($\tau_* = 30$) and a high
ionization parameter model ($\tau_* = 3000$). The structure of these
models for different values of $\Mach\Max$ is shown in
Figure~\ref{fig:wdz}. In the low-\IonPar{} model, the direct global
effects of advection are expected to be significant and indeed the
thickness of the ionized slab is reduced by 50\% as $\Mach\Max$
approaches unity. In the high-\IonPar{} model (more representative of
typical \hii{} regions), the direct global effects of advection are
expected to be negligible since $\lamad\simeq 3\times
10^{-3}\Mach\Max$. However, we find that even in this case the
thickness of the ionized slab varies by about 5\% as $\Mach\Max$ is
varied between 0 and 1. This is due to a pronounced peak that develops
in the electron density distribution at the ionization front for
$\Mach\Max \gtrsim 0.5$ (for a static model the electron density
declines monotonically through the front). We also find that the
ionization front becomes substantially sharper as $\Mach\Max$ is
increased, which is due to a decrease in the ionization fraction for a
given value of the optical depth to ionizing radiation. Both these
indirect effects of advection have significant effects on the
emissivity profiles of optical emission lines (see
Figure~\ref{fig:wdz-emissivity} of Appendix~\ref{sec:analytic-model}),
especially those such as \OIlam{} and \SIIbothlam{} that form close to
the ionization front. One can also calculate the spectral profiles of
emission lines from the models, as shown in
Figures~\ref{fig:wdz-lines} and~\ref{fig:wdz-linestats} of the
Appendix. Again, it is lines that form in the partially ionized zone
that show the most interesting behavior. These may show RMS line
widths roughly equal to the sound speed and significant velocity
offsets, both due to the gas acceleration in the ionization front. The
derived widths are roughly four times the thermal width of lines
emitted by light metals.

Although this analytic model has provided insight into some of the
effects of advection, it is obviously deficient in many respects. Many
physical processes have been ignored and in particular the use of a
fixed temperature profile $T(x)$ does not allow for the fact that
$T(x)$ itself may be affected by the advected flow.

\section{Adding Dynamics to Cloudy}
\label{sec:adding-dynam-cloudy}

In order to study the structure of advective ionization fronts in
greater detail, we need to include a wide range of additional physics.
This could be done in a variety of ways, for example by integrating
through the steady state equations or by including source terms in a
time-dependent hydrodynamic simulation.  In each case, an implicit
treatment of the source terms is necessary, as the many physical
processes with timescales shorter than the dynamical timescale 
lead to the problem being very stiff.

In the present work, we have chosen to adapt the photoionization code,
Cloudy, which already includes a comprehensive treatment of the
physical source terms.  Cloudy, in common with other traditional
plasma codes, searches for equilibrium solutions of the ionization
equations.  In order to treat steady advective flows, we have included
advective source and sink terms in the equilibrium balance equations.
The effect of this is that the equilibrium search phase now in fact
determines the implicit solution of the advective equations, and so
treats short-timescale processes in a stable manner.

In this section, we present the basic equations and outline the 
methods which we use to solve them.

\subsection{Equations} 

Cloudy takes into account the conservation equations for each species
and also the heating and cooling balance under the simplifying
assumptions of constant density or constant pressure. This procedure
can be summarized as balancing source and sink terms for the
ionization and energy equations. For the ionization equation in the
static case this can be expressed as
\begin{equation}
  \frac{d n_i}{d t} = G_i + \sum_{j\ne i} R_{j\rightarrow 
    i} n_j - n_i \left( S_i + \sum_{j\ne i} R_{i\rightarrow j} \right)
  = 0 ,
\label{e:ionization}
\end{equation}
where $d n_i/ d t$ is the rate of change of the volume
density of a particular ionization state, which in equilibrium is
equal to zero. The $ R_{i\rightarrow j}$ are the rates for ionization
(where $j$ is a higher state than $i$) and recombination (where $j$ is
a lower state than $i$). $G_i$ and $S_i$ cater for processes not
included within the ionization ladder, and are respectively the source
of ions from such processes and the sink rate into them.  A detailed
discussion of the solution method for the ionization
networks in the equilibrium case is given in
Appendix~\ref{sec:impl-deta}.

The general Cloudy solution method works by a series of nested
iterations. The innermost loop is the ionization network, external to
this is the electron density iteration, which enforces charge
neutrality, then the temperature loop, which enforces thermal balance.
Finally, an optional outermost iteration loop varies the density to
achieve pressure (or more generally momentum flux) balance. The whole
system is iterated until convergence within a given tolerance.

Once dynamics is included, the continuity and momentum equations must
be added to the set of equations to be solved, kinetic and internal
energy transport and pressure work must be taken into account, and
advection terms must be added to the ionization balance equations. For
example, for a plane-parallel steady-state flow (the simplest case,
but one that is applicable to blister \hii{} regions), the equations
to solve in flux conservative form are
\begin{equation}
  \frac{\partial}{\partial x}(\rho u) = 0 ,
\label{e:continuity}
\end{equation}
\begin{equation}
\frac{\partial}{\partial x}(p + \rho u^2) = \rho a ,
\label{e:momentum}
\end{equation}
\begin{equation}
  \frac{\partial}{\partial x}\left[\rho u \left(w + \frac{1}{2} u^2
    \right) \right] = H - C ,
\label{e:energy}
\end{equation}
\begin{equation}
  \frac{\partial}{\partial x}(n_i u) = G_i + \sum_{j\ne i} R_{j\rightarrow 
    i} n_j - n_i \left( S_i + \sum_{j\ne i} R_{i\rightarrow j} \right) .
\label{e:ad_ioniz}
\end{equation}
Here, $a$ is an acceleration e.g., gravity or radiation driving, $w$
is the specific enthalpy $w = \varepsilon + p/\rho$, where
$\varepsilon$ is the specific internal energy, and $H - C$ is heating
minus cooling.  Here the specific internal energy includes only the
thermal energy of translation, so $\varepsilon = 3/2(p/\rho)$, as
transfers from other physical energy components (ionization energy,
binding energy, vibration and rotation energy of molecules, etc.) are
treated as explicit heating and cooling terms in the underlying
thermal balance scheme.

The advection terms have the general form of $\nabla \cdot (n_i v)$
(for steady state), where $v$ is the advection velocity. This can be
written as
\begin{equation}
  \nabla \cdot (n_iv) = nv\cdot \nabla(n_i/n) . 
\end{equation}

\subsection{Differencing Scheme}
Although it is possible to solve the ionization equations in an
explicitly time-dependent way, this is not the best way to proceed.
The photoionization terms in the steady-state solution will often have
very short timescales, so stability constraints would limit the time
step to this short photoionization timescale, and hence cause an
extremely slow convergence of the iterative scheme.  Instead, we take
advantage of the current algorithm used by Cloudy and difference the
equations implicitly.  Such an implicit scheme has the advantage that
the timestep is not limited to the shortest ionization or
recombination time, which is clearly unsatisfactory for an
astrophysical system in which the dynamical timescales are general
much longer than the ionization or recombination timescales.

At iteration $m$, the advection terms may be approximated based on the
value in the present zone and an upstream value in the previous full
iteration as
\begin{equation}
  {d\over dz} x_i \to {x^m_i(z) - x^{m-1}_i(z-\Delta z)\over \Delta z},
\end{equation} where $x^m_i(z)$ is the value of $x_i = n_i/n$ at
position $z$ at the $m$th iteration of the scheme, and $\Delta z$ is
an adjustable \emph{advection length}.  For the first iteration no
upstream values are available so no advection terms are included in
the equations.  It is useful to define the look-back operator
\begin{equation}
  L_{\Delta z}[x_i^m(z)] = x_i^{m-1}(z-\Delta z)
\end{equation}
for values, such as $x_i$, given per unit material.  Values specified per unit
volume need to be scaled to a conserved variable before the look back is
applied, so that
\begin{equation}
  L_{\Delta z}[n_i^m(x)] = n^m x_i^{m-1}(z-\Delta z).
\end{equation}
This may be thought of as a first-order Lagrange-remap solution
for the advection equation.

For the scheme discussed in Appendix~\ref{sec:impl-deta} for the
ionization ladder, the advective terms may then be included
simply as an additional source term,
\begin{equation}
G_i = n v \frac{ L_{\Delta z}[x_i^m(z)] }{ \Delta z },
\end{equation}
and sink rate
\begin{equation}
S_i = {v\over \Delta z}
\end{equation}
in the linearized form of the equations, and iterated to find the
non-linear solution as in the time-steady case.

The energy balance equation may be treated in a similar manner.  Using
the mass conservation equation (eq.~\ref{e:continuity}), we have from
equation~(\ref{e:energy}) 
\begin{equation}
\rho v.\nabla \left(w+{1\over 2} v^2\right) = H-C,
\end{equation}
which may be differenced as
\begin{equation}
\rho v {\left(w+{1\over 2} v^2\right)-L_{\Delta z}\left(w+{1\over 2} v^2\right)
\over\Delta z} = H-C.
\end{equation}
The terms on the left side of this equation may then be treated as
additional heating and cooling terms in the temperature solver.

The continuity and momentum equations are more easily dealt with. The
continuity equation is taken into consideration simply by eliminating
the velocity $v$ in terms of $\rho$ in all equations using
the substitution
\begin{equation}
  \rho v r^d = \mathrm{constant} ,
\end{equation}
which comes from integrating the general form of
equation~\ref{e:continuity} where $d = 0, 1, 2$ indicates plane
parallel, cylindrical or spherical geometry, respectively. The initial
condition is given at the illuminated face. 
  
The dynamical pressure, which appears in the momentum equation, is
taken into account by adding the ram pressure term $\rho v^2$ to the
total pressure.

\paragraph{Varying the Advection Length}

The advection length, $\Delta z$, in this scheme determines the manner
in which different processes are treated.  The differencing we have
chosen has the effect that processes far more rapid than $\Delta z /
v$ are treated as in static equilibrium, while slower processes are
followed exactly.  The correct steady-state solution is found in the
limit $\Delta z \to 0$, but the smaller the advection length chosen,
the longer the system will take to reach an equilibrium state.  The
natural procedure is then to use a first iteration solution by
ignoring the advection terms, and then gradually decrease $\Delta z$
until $\Delta z \sim\Delta z\Sub{grid}$, where $\Delta z\Sub{grid}$ is
the size of a spatial zone in the simulation, at which stage the
treatment of advection will be as accurate as that of photoionization.

In order to track the convergence of the models and to determine when
to reduce the advective timestep, we monitor the behavior of two error
norms, $\epsilon_1$ and $\epsilon_2$. Both are calculated as the
squared norm over all zones, $z$, and ionic/molecular species, $i$. 
The first of these norms is the \emph{convergence error}, defined as 
\begin{equation}
\epsilon_1 = \left\Vert n_i^{m} - n_i^{m-1}\over 
  \Delta z / v \right\Vert_{z,i},
\end{equation}
which measures the difference between the model solutions for the last
two iterations. The second is the \emph{discretization error}, defined
as
\begin{equation}
\epsilon_2 = \left\Vert {n_i - L_{\Delta z}(n_i)\over \Delta z / v} -
  {n_i - L_{(\Delta z/2)}(n_i)\over \Delta z / 2v } \right\Vert_{z,i},
\end{equation}
which measures the accuracy of the present estimate of the advective
gradients in the solution compared to an estimate with half the
advection length.

If $\epsilon_1\ll\epsilon_2$, then the solution is converged with the
present $\Delta z$, while if the errors in the Lagrangian estimate of
the gradient of the value are still significant, then the timestep
should be decreased.  Cutting $\Delta z$ when $\epsilon_1^2 <
0.1\epsilon_2^2$ produces substantial improvements in the rate of
convergence of the advective solutions.  However, it can still take a
substantial number of iterations to reach equilibrium for large
advection velocities. 

\begin{figure}
  \includegraphics[width=\linewidth]{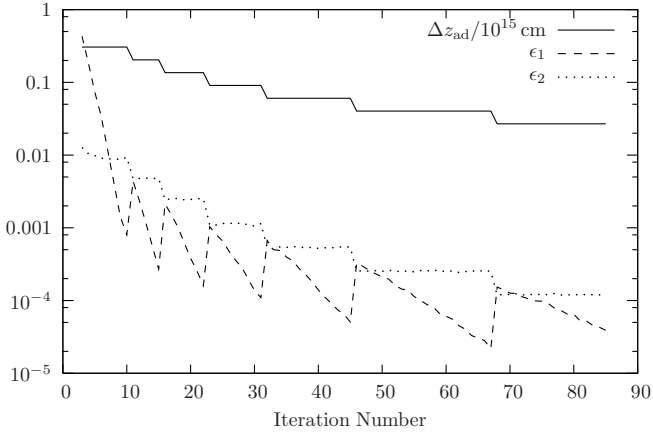}
  \caption{Convergence behavior of an example model as a function of
    iteration number, $m$. Solid line: advection length $\Delta z$
    (units of $\unit{10^{15}}{\centi\meter}$); dashed line: convergence
    error, $\epsilon_1$; dotted line: discretization error,
    $\epsilon_2$.}
  \label{fig:converge}
\end{figure}

An example of the way in which $\Delta z$, $\epsilon_1$, and
$\epsilon_2$ vary during a model calculation is given in
Figure~\ref{fig:converge} (which corresponds to model ZL009 discussed
in the following section). It can be seen that while $\Delta
z\Sub{ad}$ remains constant, the convergence error, $\epsilon_1$,
decreases with each iteration, while $\epsilon_2$ hardly changes after
the first iteration for a given $\Delta z$. When
$\epsilon_1/\epsilon_2$ falls to a low enough value, then all physical
processes that occur on timescales longer than $\Delta t = \Delta z /
v$ have converged, so the advection length can be reduced.  This has
the effect of lowering $\epsilon_2$ but also temporarily increases
$\epsilon_1$ due to the release of shorter-timescale processes from
strict local equilibrium, so several iterations must be carried out at
the new value of $\Delta z$. This procedure is continued until
$\epsilon_2$ has fallen to a sufficiently low value.

\section{Results}
\label{sec:results}

In this section we present results for selected advective ionization
fronts calculated using Cloudy, all using a plane-parallel slab
geometry.  The principal input parameters for the models are the
hydrogen number density, $n_0$, gas velocity, $u_0$, hydrogen-ionizing
photon flux, $F_0$, all specified at the illuminated face.  The
spectral distribution of the incident radiation field was assumed to
be a black body with effective temperature, $T\Sub{eff}$.  All these
parameters are shown in Table~\ref{tab:params} for the three models
presented here. The gas-phase elemental abundances for all the models
were set at the standard ISM values \citep{1996ApJ...468L.115B} and
Orion-type silicate and graphite grains were included
\citep{1991ApJ...374..580B}. Since this paper is concerned with the
effects of advection on the ionization front, all molecules were
turned off and the integration was stopped when the electron fraction
fell below $10^{-3}$. The inclusion of molecular processes in the PDR
will be described in a following paper. The models also include an
approximate treatment of a tangled magnetic field (see
Appendix~\ref{sec:magnetic-field}), characterized by the field
strength at the illuminated face, $B_0$.

One further physical process was disabled in these models: the
radiative force due to the absorption of stellar continuum radiation
(principally by dust grains). This was done for purely pragmatic
reasons, since the inclusion of this process for high ionization
parameter models makes it very difficult to set the approximate
desired conditions at the ionization front by varying conditions at
the illuminated face. Models that do include this process are
discussed further in section~\ref{sec:veloc-ioniz-corr}. 


The first two models, ZL009 and ZH007, are weak-D fronts with low and
high ionization parameter, respectively, with parameters similar to
the toy models discussed in Appendix~\ref{sec:analytic-model}. The
velocity at the ionized face, $u_0$, was set somewhat below the
isothermal sound speed in order to avoid the possibility of gas
passing through a sonic point during an intermediate iteration
(transonic fronts will be considered in a following paper). The third
model, ZH050, is a weak-R front in which the gas velocity relative to
the front is supersonic throughout. Such R-type fronts are likely to
be transient and thus of limited observational
significance. Nevertheless, this model is useful since it provides a
stringent test of our simulations in the limit of high advective
velocities. 

As well as the advective models, we also calculate equivalent static
models for each of the three cases considered, which have constant
pressure for comparison with the weak-D fronts or constant density for
comparison with the weak-R front. 

\begin{table}[!tp]\centering
  \caption{Model Parameters\label{tab:params}}
  \begin{tabular}{l@{\hspace{4\tabcolsep}}rrr}
    \toprule
    Parameter  & ZL009 & ZH007 & ZH050  \\ \midrule
    $\log ( n_0 /\mathrm{\rpcubic{\centi\meter}} )$ &  $3.5$ & $4.0$ & $4.0$  \\
    $u_0 / \mathrm{\kilo\meter\usk\reciprocal\second}$  & $-9.0$ & $-7.0$ & $-50.0$  \\
    $\log ( F_0 /\mathrm{\rpsquare{\centi\meter}} )$& $9.5$ & $13.0$ & $13.0$  \\ 
    $\log ( T\Sub{eff} / \mathrm{\kelvin} )$& $4.6$ & $5.0$ & $4.6$  \\
    $\log ( B_0 / \mathrm{Gauss} )$& none & $-4.0$ & $-4.0$  \\[\medskipamount] 
    $\log \IonPar_0$ & $-4.5$ & $-1.5$ & $-1.5$ \\
    $\tau_* $ & 9.9 & 6800 & 6800 \\
    $\Mach\Max$ & $ 0.84$ & $0.73$ & $4.01$ \\ 
    $\xiad$ & 8.1 & 13.8 & 13.8 \\
    $\lamad$ & 2.2 & 0.0015 & 0.008\\
    \bottomrule
  \end{tabular}
\end{table}

\subsection{Low Ionization Parameter, Weak-D}
\label{sec:metals-no-molecules}

%
%
\begin{figure}\centering
  \includegraphics
  [width=\linewidth]
  {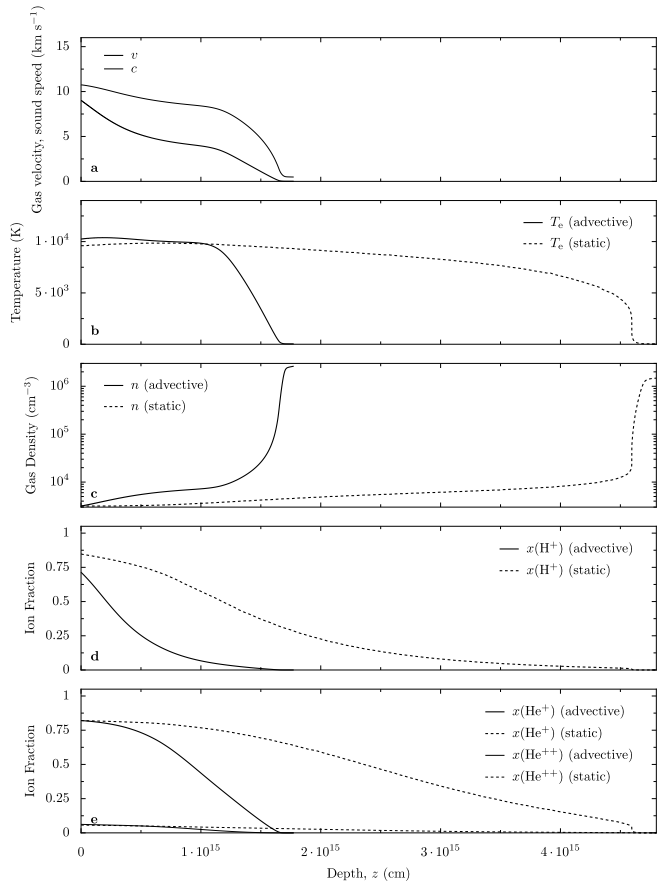}
  \caption{Structure of model ZL009 as a function of depth from the
    illuminated face. (\textit{a}) Velocity and isothermal sound
    speed. (\textit{b}) Gas temperature. (\textit{c}) Number density
    of hydrogen nucleons.  (\textit{d}) Hydrogen ionization fraction.
    (\textit{e}) Helium ionization fractions. Panels
    \textit{b}--\textit{e} show the advective model results (solid
    line) and the results from an equivalent static constant pressure
    model (dashed line).}
  \label{fig:ZL009a}
\end{figure}
\begin{figure}\centering
  \includegraphics
  [width=\linewidth]
  {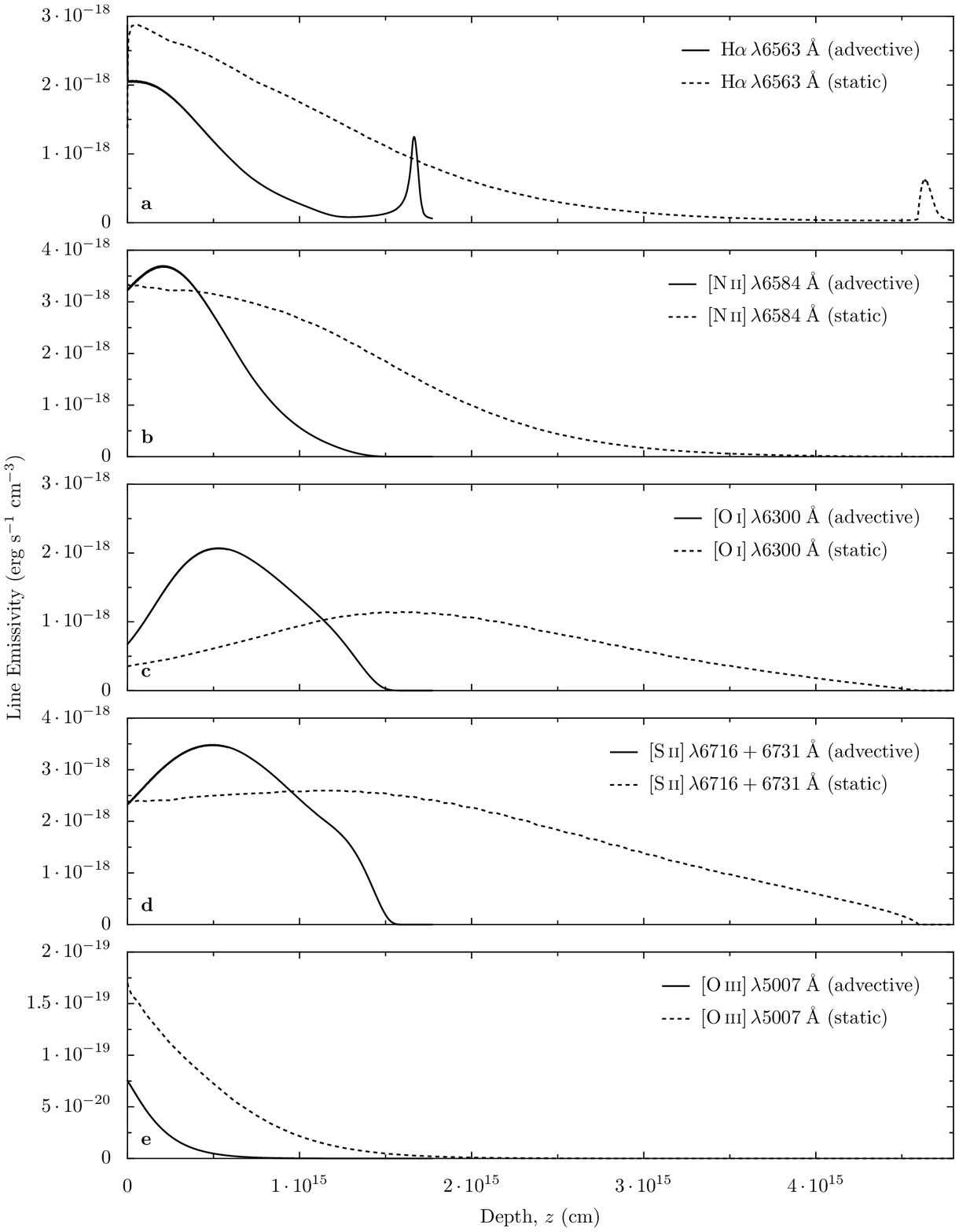}
  \caption{Emissivity structure of model ZL009. 
    As Figure~\ref{fig:ZL009a} but showing volume emissivity of
    (\textit{a})~\Halam{}, (\textit{b})~\NIIlam{},
    (\textit{c})~\OIlam{}, (\textit{d})~\SIIbothlam{}, and
    (\textit{e})~\OIIIlam.  }
  \label{fig:ZL009c}
\end{figure}
\begin{figure}\centering
  \includegraphics
  [width=\linewidth]
  {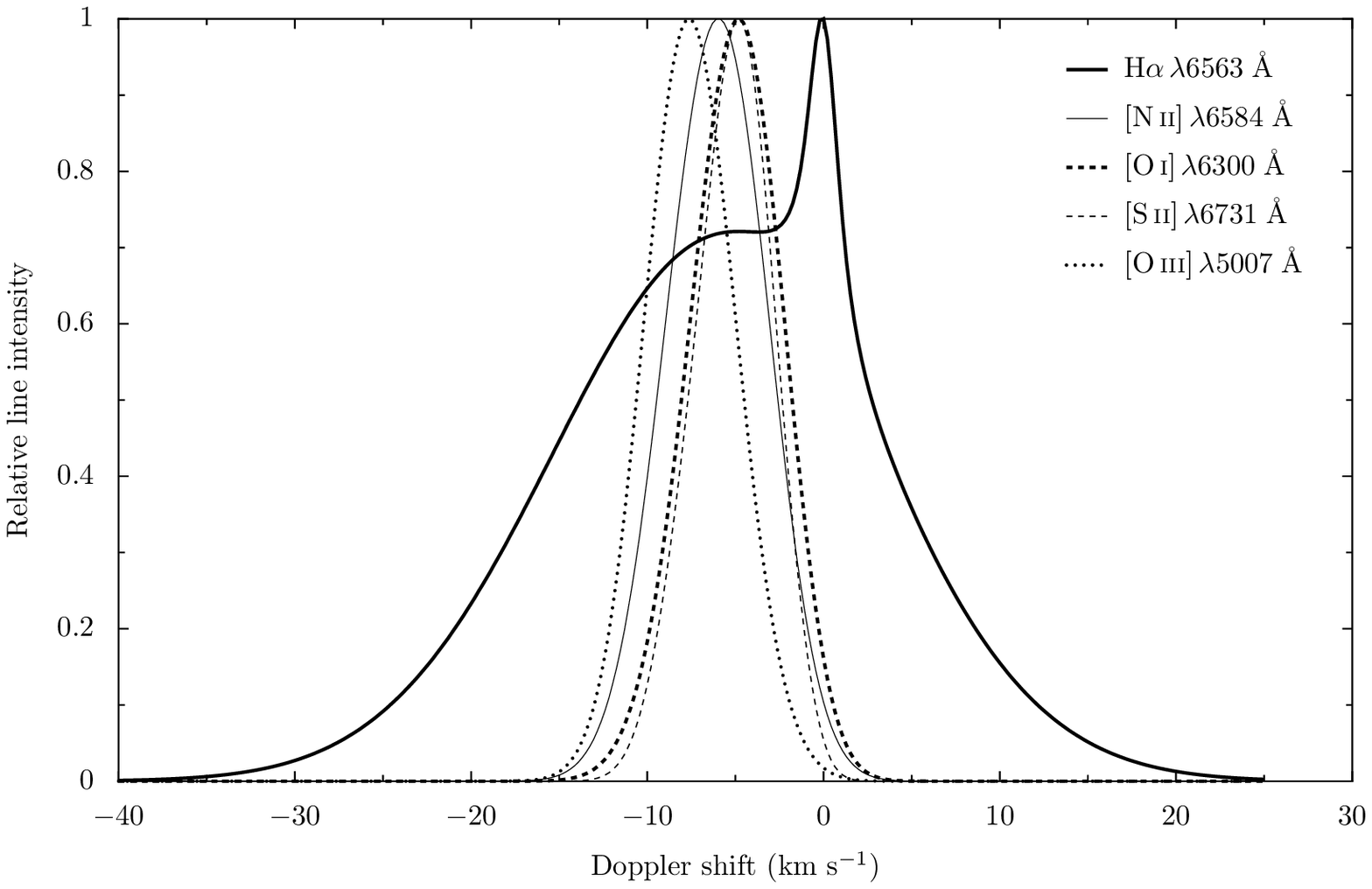}
  \caption{Face-on emission line profiles of model ZL009
    including thermal broadening: \Halam{} (thick solid line),
    \NIIlam{} (thin solid line), \OIlam{} thick dashed line),
    \SIIlam{} (thin dashed line), and \OIIIlam{} (thick dotted line).
    All lines are normalized to their peak intensities.  }
  \label{fig:ZL009d}
\end{figure}

This model, ZL009, has physical parameters that are inspired by those
of cometary globules in planetary nebulae such as the Helix, although
the geometry is plane-parallel rather than spherically divergent. The
ionization parameter of the model is very low ($\IonPar = 3.33\times
10^{-5}$), which accentuates the global effects of advection and also
leads to the ionization front thickness being comparable to the
Strömgren thickness of the ionized layer.

As can be seen from Figure~\ref{fig:ZL009a}, the advection has large
effects on the model structure as would be expected given the large
value of $\lamad$. The depth of the ionized region is reduced by more
than a factor of two with respect to the static model, with
concomitant reductions in the brightness of the hydrogen recombination
lines (see Table~\ref{tab:ZL009-lines}). The collisional lines of
\OI{}, \SII{}, and \NII{} are also reduced in intensity, albeit to a
lesser degree (\OIII{} emission from this model is negligible due to
the low ionization parameter). Interestingly, the intrinsic Balmer
decrement is increased in the advective model, which gives $\Ha/\Hb =
3.29$ (reddening by internal dust is not included in the line ratios
given in Table~\ref{tab:ZL009-lines}).  This is because the
temperature at low ionization fractions is significantly higher in the
advective model, which leads to a non-negligible collisional
contribution to the Balmer emission that preferentially excites
H$\alpha$.  The temperature in the more highly ionized zone is also
increased somewhat by the effects of advection.

\begin{table*}[t!]\centering
  \caption{Emission Line Properties from Weak-D Photoionization Models}
  \begin{tabular}{lr@{\qquad} rr rr c@{\qquad} rr rr }
    \toprule
    & & \multicolumn{4}{c}{Model ZL009} & \qquad & 
    \multicolumn{4}{c}{Model ZH007} \\
    \cmidrule{3-6}\cmidrule{8-11} 
    & & \multicolumn{2}{c}{Line Intensity} & & & & 
    \multicolumn{2}{c}{Line Intensity} \\
    Line & $\lambda$ & 
    Static & Advect & $\bar{v}$ & $\Delta v$ & & 
    Static & Advect & $\bar{v}$ & $\Delta v$ \\
    \midrule
    \Ha &   $6563$ & $2.95$ & $3.29$ & $-5.89$ & $21.6$ & & $2.92$ & $2.92$ & $-6.46$ & $20.0$ \\
    \OI &   $6300$ & $2.40$ & $4.91$ & $-5.04$ & $6.3$  & & $3.5\,(-3)$ & $2.8\,(-3)$ & $-6.99$ & $6.4$  \\
    \NII &  $6584$ & $3.97$ & $6.23$ & $-6.07$ & $6.9$  & & $0.42$ & $0.43$ & $-7.56$ & $6.0$  \\
    \SII &  $6731$ & $3.52$ & $5.33$ & $-5.17$ & $5.3$  & & $0.04$ & $0.03$ & $-7.29$ & $4.3$  \\
    \OIII & $5007$ & $0.05$ & $0.03$ & $-7.50$ & $6.1$  & & $5.21$ & $5.21$ & $-6.32$ & $5.0$  \\[\smallskipamount] 
    \Hb & $4861$  & $-2.868$ & $-3.382$ & & & & $0.435$ & $0.440$ \\
    \bottomrule
  \end{tabular}
  \label{tab:ZL009-lines}
  \parbox{0.51\linewidth}{
  \tablecomments{{}Line intensities are all calculated for a face-on
    orientation and are given relative to \Hb{} except for \Hb{}
    itself, which is $\log(\mathrm{intensity})$ in units of
    erg\,cm$^{-2}$\,s$^{-1}$. }
  }
\end{table*}

Also listed in Table~\ref{tab:ZL009-lines} are the mean velocities and
full-width-half-maxima of the different emission lines, calculated
assuming that the model is observed face on. The emission line
profiles are illustrated in Figure~\ref{fig:ZL009d}. The contribution
from each computational zone to the line profile is thermally
broadened using the local temperature and the atomic weight of the
emitting species. As a result, the \Ha{} line is significantly broader
than the collisionaly excited metal lines. The \NII{} line is blue
shifted with respect to \SII{} and \OI{}, which is due to the
acceleration of gas through the ionization front, as can be seen in
Figure~\ref{fig:ZL009a}\textit{a}. The \Ha{} line has two components:
a broad blue-shifted component due to emission from the ionized flow,
and a narrow component at zero velocity, caused by a subsidiary peak
in the electron density that occurs at low ionization fractions where
the temperature changes sharply.

For the \SII{} line, the thermal broadening and the gas acceleration
both contribute in equal amounts to the predicted line width. For the
lighter metal lines, the acceleration broadening remains roughly the
same but the thermal width is increased somewhat. For \Ha{}, the
thermal broadening dominates.

In this model, the gas velocity and Mach number increase monotonically
as the gas flows from the neutral to the ionized side, reaching a
maximum Mach number of $\Mach\Max = 0.84$ at the illuminated face.  As
a result, emission lines from more highly ionized species are more
blue-shifted (see Table~\ref{tab:ZL009-lines}). However, the effect is
slight with only \unit{2.5}{\kilo\meter\usk\reciprocal\second}
velocity difference between \OI{} and \OIII{}.

\subsection{High Ionization Parameter, Weak-D}
\label{sec:metals-highZ}

\begin{figure}\centering
  \includegraphics
  [width=\linewidth]
  {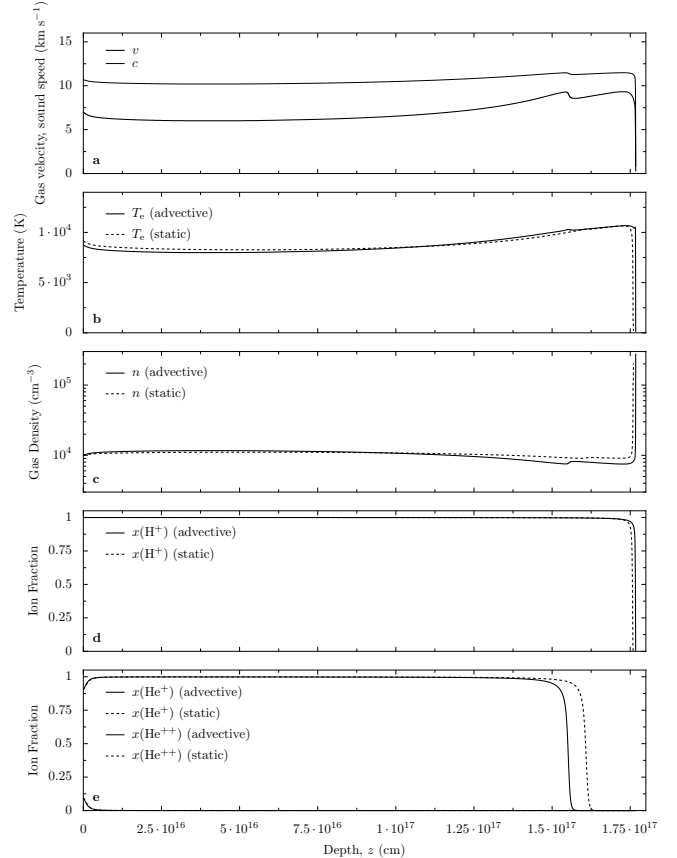}
  \caption{Structure of model ZH007 as a function of depth from the
    illuminated face. All panels as in Figure~\ref{fig:ZL009a}.}
  \label{fig:ZH007a}
\end{figure}
\begin{figure}\centering
  \includegraphics
  [width=\linewidth]
  {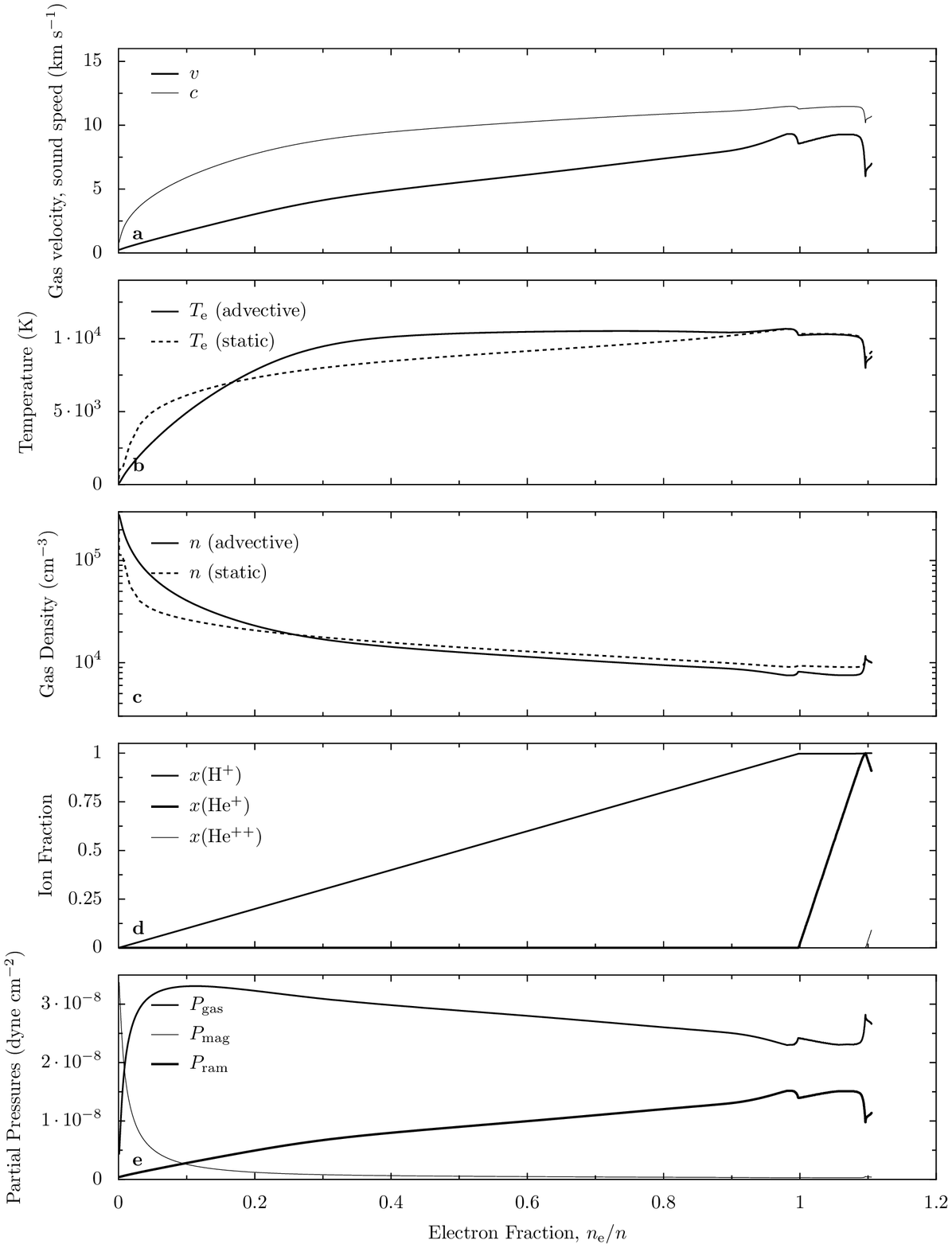}
  \caption{Structure of model ZH007 as a function of electron
    fraction, $n\Sub{e}/n$. Panels (\textit{a})--(\textit{c}) as in
    Figure~\ref{fig:ZH007a}. (\textit{d}) Ionization fractions of
    H$^+$ (medium weight line), He$^+$ (thick line), and He$^{2+}$
    (thin line). (\textit{e}) Partial contributions to the
    total pressure: thermal gas pressure (medium weight line), magnetic
    pressure (thin line), and ram pressure (thick line).  }
  \label{fig:ZH007b}
\end{figure}
\begin{figure}\centering
  \includegraphics
  [width=\linewidth]
  {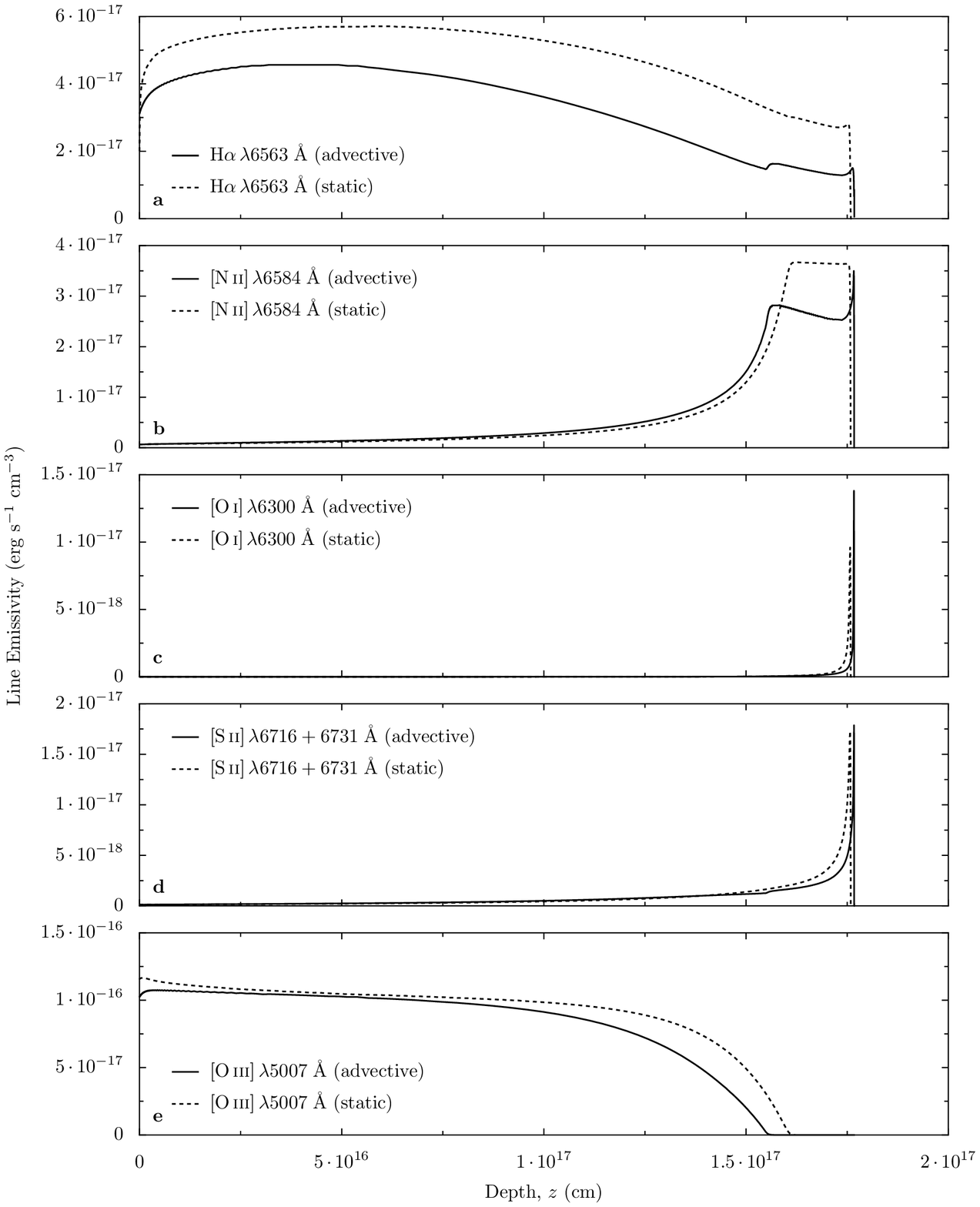}
  \caption{Emissivity structure of model ZH007 in units of
    \Emissivity{2\times10^{-16}}). Panels as in
    Figure~\ref{fig:ZL009c}.  }
  \label{fig:ZH007c}
\end{figure}
\begin{figure}
  \includegraphics
  [width=\linewidth]
  {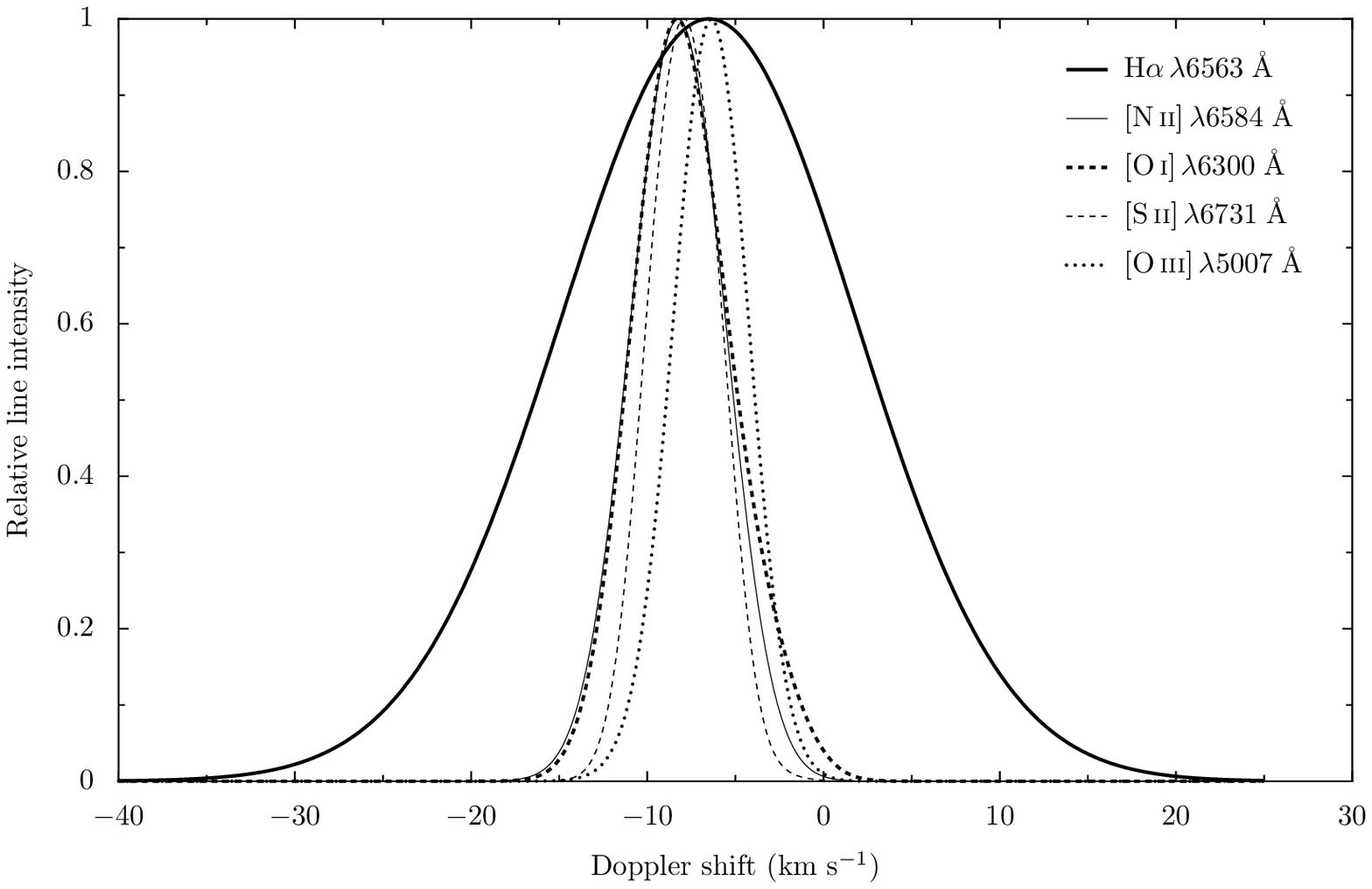}
  \caption{Face-on emission line profiles of model ZH007
    including thermal broadening.  \Halam{} (solid lines), \NIIlam{}
    (dashed lines), \OIlam{} (dotted lines), and \SIIlam{} (dot-dashed
    lines).  Vertical axis units are arbitrary.  }
  \label{fig:ZH007d}
\end{figure}

This model, ZH007, has physical parameters (see
Table~\ref{tab:params}) inspired by the central region of the Orion
nebula \citep{1991ApJ...374..580B}. The density is only slightly
higher than in ZL009 but the ionizing flux is much higher, resulting
in a far higher ionization parameter ($\IonPar = 3.3\times10^{-2}$).
The ionization front is much thinner than the depth of the fully
ionized slab. The local advection parameter, $\xiad$ (see
Section~\ref{sec:advect-ioniz-fronts} and
Appendix~\ref{sec:analytic-model}), is somewhat higher than in the
previous model due to the hotter gas temperature and softer ionizing
spectrum but the much higher value of $\tau_*$ that accompanies the
higher ionization parameter means that the global advection parameter,
$\lamad$, is very small.

The structure of the advective model as a function of depth into the
slab from the illuminated face is shown by solid lines in
Figure~\ref{fig:ZH007a}.  For some panels, an equivalent
constant-pressure, static model is also shown (dashed lines). The
indirect effects of advection are much greater than the direct loss of
0.1\% of the incident ionizing flux due to the ionization of fresh
gas. The largest effect is a roughly 3\% increase in the mean density
in the ionized zone (the densities at the illuminated faces are set
equal in the static and advective models) due to the varying
importance of ram pressure as the velocity varies through the slab.
Due to the difference in density dependence of recombination and dust
absorption, this leads to a slight decrease in the ionized column
density together with an increase in the emission measure, caused by a
reduction in the fraction of ionizing photons that are absorbed by
dust grains.

Figure~\ref{fig:ZH007b} shows some of the same quantities as in
Figure~\ref{fig:ZH007a} but this time plotted against the electron
fraction, $x\Sub{e} = n\Sub{e}/n$. This effectively `zooms in' on the
ionization front transition itself, allowing one to appreciate details
of the structure that are not apparent in the plots against depth.
Note that the more neutral gas is on the left in
Figure~\ref{fig:ZH007b}, whereas it was on the right in
Figure~\ref{fig:ZH007a}. In Figure~\ref{fig:ZH007b}\textit{b}, which
shows the temperature profiles, it can be seen that the initial
heating of the gas as it is ionized is more gradual in the advective
model than in the static model due to the photo-heating timescale
being longer than the dynamic timescale. However, once the ionization
fraction exceeds about 20\% the photo-heating rate exceeds the static
equilibrium model because the 
neutral fraction at a given value of the ionizing flux is higher than in the static case. 
This produces a characteristic overheating, which is often seen in
advective fronts.  In the current case it is relatively modest,
producing an extended $\unit{10^4}{\kelvin}$ plateau for $x\Sub{e} =
0.3$--0.9. For $x\Sub{e} > 0.9$ the gas approaches static thermal
equilibrium again and the two curves coincide with temperature
variations determined by radiation hardening and the varying
importance of the diffuse field.  Figure~\ref{fig:ZH007b}\textit{c}
shows the variation in the gas density and it can be seen that the
density on the far neutral side is significantly higher in the
advective model. This is a result of the lack of ram pressure on the
neutral side (see Figure~\ref{fig:ZH007b}\textit{e}), which means the
thermal pressure must increase to compensate. The magnetic field at
the illuminated face in this model was set to be \unit{0.1}{\milli G},
implying a negligible contribution to the total pressure in the
ionized gas.  However, on the far neutral side, where the density is
higher the field is much larger (assumed to grow with compression as
$B\propto \rho^{2/3}$, see Appendix~\ref{sec:magnetic-field}), so that
the magnetic pressure becomes appreciable, limiting the density in the
cold gas. The inferred $B$ in the neutral gas is similar to observed
values in the neutral veil of Orion \citep{1989ApJ...337..342T}.
%

The line emissivities as a function of depth for the ZH007 model are
shown in Figure~\ref{fig:ZH007c}. %
The main change between the static and advective
models is the sharp peak at the ionization front that is seen in the
emissivity of lines from singly-ionized species such as \NII{}. This
peak is due to the electron density peak seen in Figure~\ref{fig:wdz}
and discussed in Appendix~\ref{sec:analytic-model}.

The integrated emergent emission line spectrum
(Table~\ref{tab:ZL009-lines}) is barely affected by the advection.  As
a result of the increase in emission measure discussed above, although
the ionization front moves appreciably inward
(Figure~\ref{fig:ZH007a}\textit{d}), the Balmer line flux is actually
slightly \emph{higher} in the advective model than in the static
model.  On the other hand, lines that form in the ionization front
itself, such as \SII{} and \OI{}, are somewhat reduced in intensity in
the advective model due to the narrowing of the ionization front by
advection (see discussion in Appendix~\ref{sec:analytic-model}).

Unlike in the the low ionization parameter model of the previous
section, in this model the velocity and Mach number do not increase
monotonically from the neutral to the ionized side. Instead, the
maximum Mach number ($\Mach\Max = 0.73$), which also corresponds to
the maximum velocity, occurs at the He$^0$/He$^+$ front. Although this
does not correspond to the maximum temperature, it is a maximum in the
sound speed due to the decrease in the mean mass per particle when He
is ionized. There is a second, slightly lower, maximum in the sound
speed, Mach number, and velocity just inside the H ionization front
where radiation hardening causes a temperature maximum. The velocity
also starts to increase again very close to the illuminated face. 

Due to this complex velocity structure, the trends of blueshift with
ionization parameter are less clear in this model, as can be seen from
Table~\ref{tab:ZL009-lines} and from Figure~\ref{fig:ZH007d}, which
shows simulated emission line profiles.

\subsection{High Ionization Parameter, Weak-R}
\label{sec:metals-highZ-weakR}

This model, ZH050, is identical to ZH007 except that the gas velocity
at the illuminated face is set to \Velocity{50}, producing a weak-R
front. The model structure as a function of depth is shown in
Figure~\ref{fig:ZH050a} and as a function of electron fraction in
Figure~\ref{fig:ZH050b}. In both cases, the advective model is now
compared with a constant-density static model as opposed to the
constant-pressure model that was used for comparison with the weak-D
models. As can be seen from Figure~\ref{fig:ZH050a}\textit{a} and
\textit{c} the velocity and density are roughly constant across the
front, as is expected in the weak-R case.  The extremal Mach number in
the front (which for R-type fonts is a minimum, see
Appendix~\ref{sec:analytic-model}) is $\Mach\Max = 4.01$, which occurs
on the ionized side of the front (see Fig.~\ref{fig:ZH050b}\textit{c})
at $x\Sub{e} \simeq 0.93$.

\subsection{Temperature Structure of the Ionization Fronts}
\label{sec:temp-struct-ioniz}

\begin{figure}\centering
  \includegraphics
  [width=\linewidth]
  {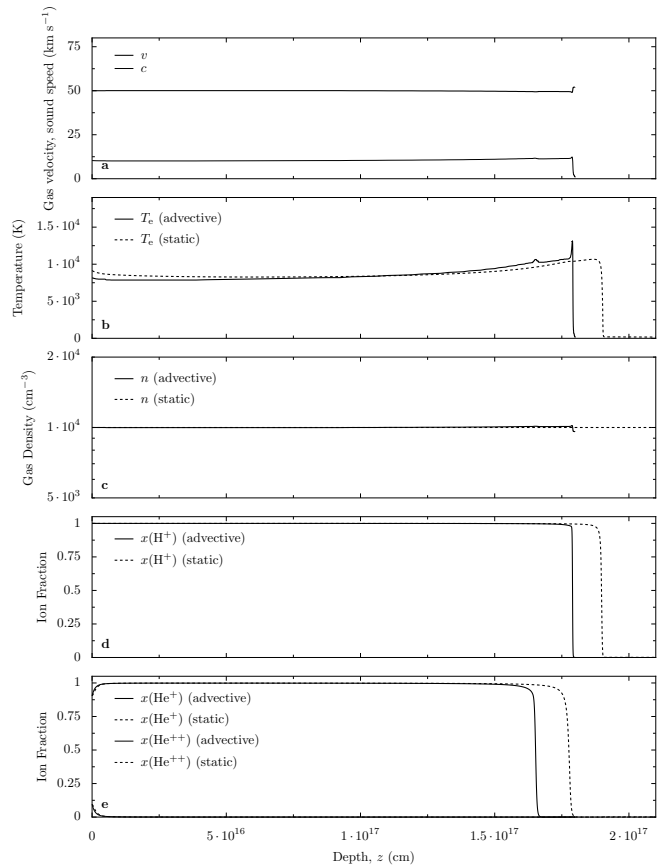}
  \caption{Structure of model ZH050 as a function of depth from the
    illuminated face. All panels as in Figure~\ref{fig:ZL009a}.}
  \label{fig:ZH050a}
\end{figure}

\begin{figure}\centering
  \includegraphics
  [width=\linewidth]
  {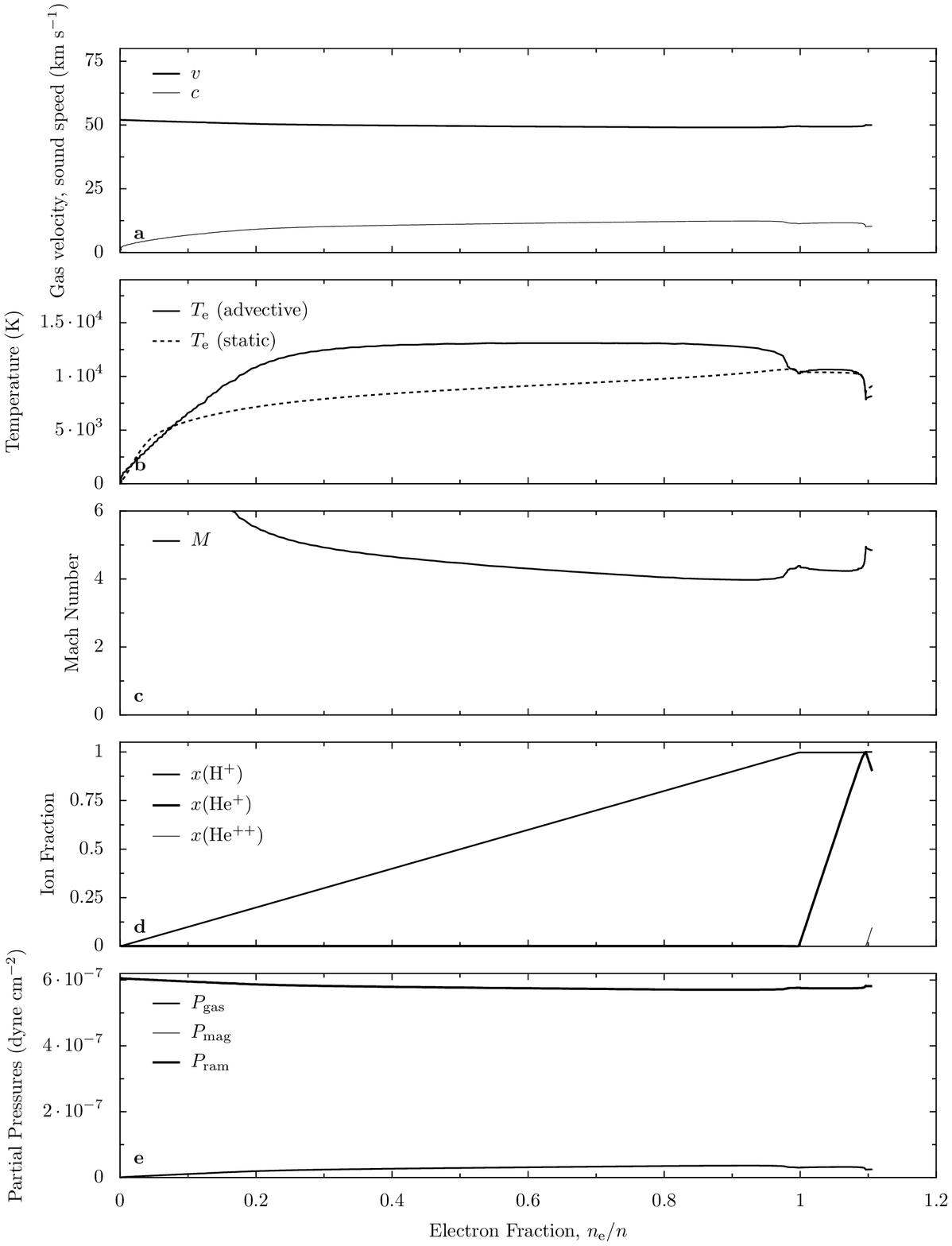}
  \caption{Structure of model ZH050 as a function of electron
    fraction, $n\Sub{e}/n$. All panels as in Figure~\ref{fig:ZH007b}
    except for (\textit{c}), which shows the isothermal Mach number.}
  \label{fig:ZH050b}
\end{figure}

Figure~\ref{fig:ZH050a}\textit{b} shows that our weak-R model has a
pronounced temperature spike at the ionization front, together with a
smaller spike at the He ionization front.
Figure~\ref{fig:ZH050b}\textit{b} shows that this is a more extreme
manifestation of the overheating in the ionization front that was seen
in the weak-D model (Fig.~\ref{fig:ZH007b}\textit{b}). However, even
in this case of a rapidly propagating front we do not find the
overheating to be very great, reaching a maximum temperature of only
\unit{13,000}{\kelvin}.  Other authors have found more pronounced
over-heating effects in dynamic ionization fronts
\citep{1998A&A...331..347R,1997A&A...325.1132M} but for different
values of the physical parameters. \citeauthor{1997A&A...325.1132M}
found temperatures as high as \unit{20,000}{\kelvin} behind R-type
fronts moving at a substantial fraction of the speed of light, which
would be difficult to model using our steady-state code.
\citeauthor{1998A&A...331..347R} studied the time-dependent evolution
of an \hii{} region after the turning-on of an O~star, finding
temperatures of order \unit{15,000}{\kelvin} behind the ionization
front soon after its transition from R to D-type. Their higher
temperatures may be due to these authors including fewer cooling
processes than are included in Cloudy \citep[for further discussion,
see][ and references therein]{2001MNRAS.325..293W}. The greater width
of the temperature spike in the \citeauthor{1998A&A...331..347R}
models is due to the fact that they were considering lower densities,
so the gas moves further from the ionization front in the time it
takes to cool to the equilibrium temperature.




\section{Discussion}
\label{sec:discussion}

In this section, we look for evidence of advective effects in one of
the closest and best-studied \hii{} regions, the Orion nebula. We
concentrate on the clearest signatures of advection to emerge from our
simulations: the electron density spike and the ionization-resolved
kinematics.

\subsection{Electron Density Structure of the Orion Bar}
\label{sec:electr-dens-struct}

\begin{figure*}
  \includegraphics[width=\linewidth]{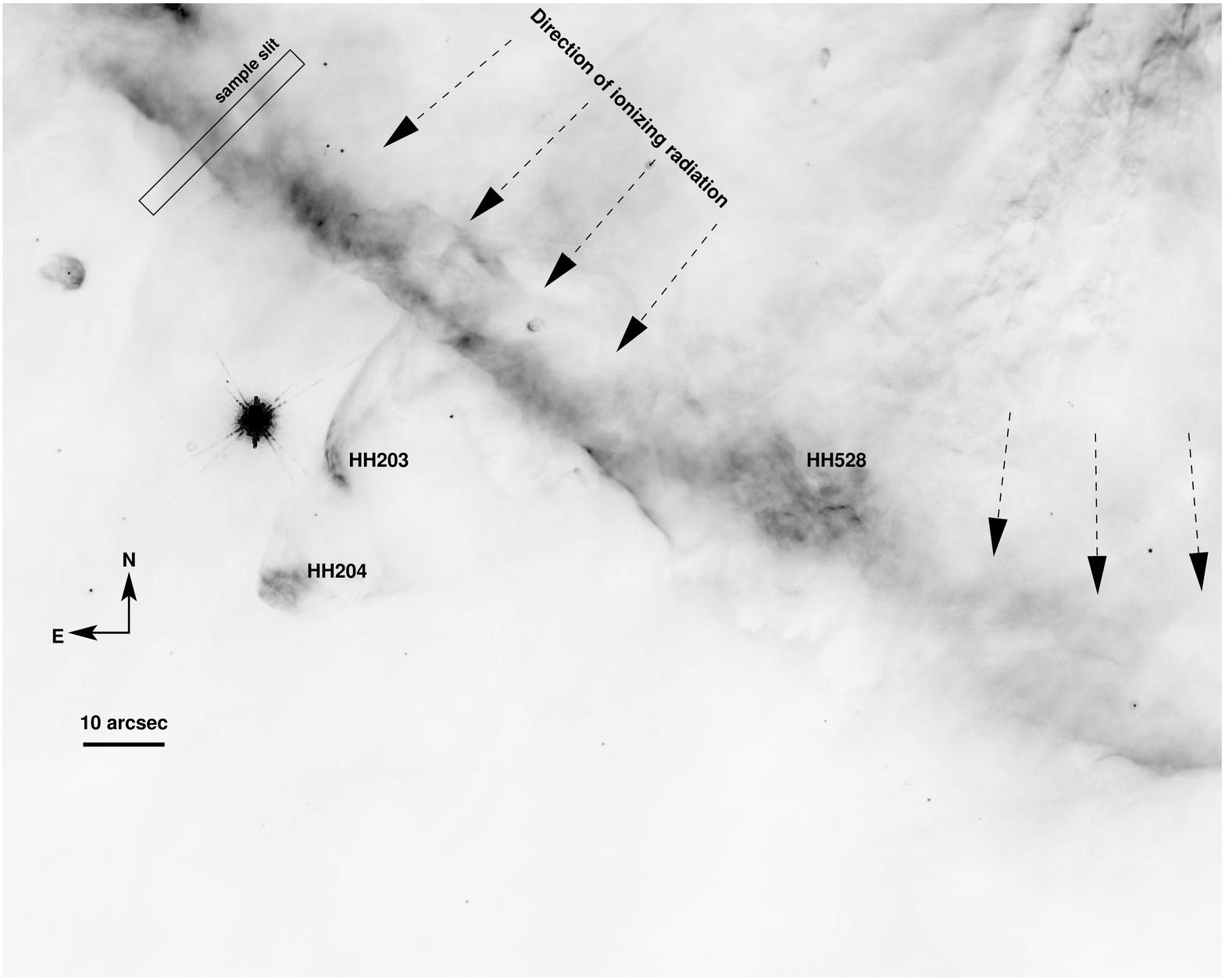}
  \caption{Negative \textit{HST} WFPC2 image of the Orion bar in the
    \NIIlam{} line. Image dimensions are $150 \times 120''$.}
  \label{fig:bar-nii}
\end{figure*}

\begin{figure}
  \includegraphics[width=\linewidth]{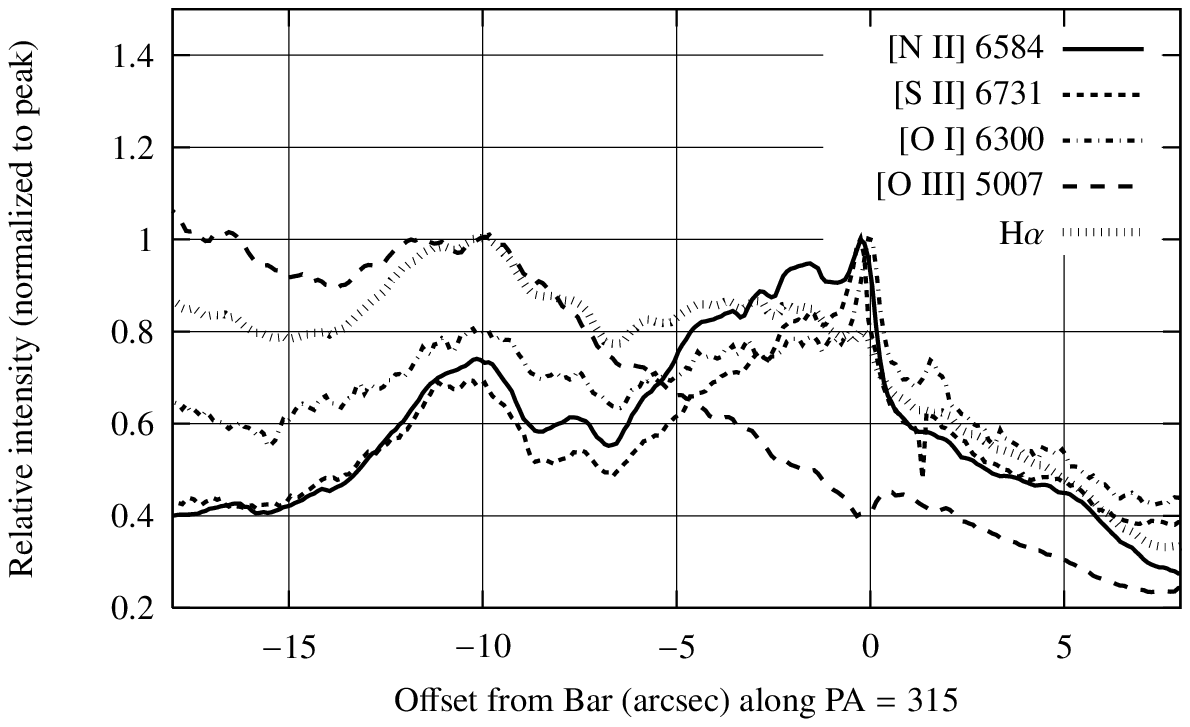}
  \caption{ \textit{HST} WFPC2 Spatial intensity profiles of various
    emission lines along a line perpendicular to the Orion bright bar.
    The ionizing source lies of the graph at an offset of $\simeq
    -120''$.}
  \label{fig:bar-profiles}
\end{figure}

One firm prediction of the advective model that differs from the
static case is the existence of a sharp electron density peak at the
position of the ionization front. This peak manifests itself most
clearly in the emission of the \NIIlam{} line (see
Figure~\ref{fig:ZH007c}), producing a narrow emissivity peak at the
edge of the broader peak that comes from the neutral Helium zone.  In
Figure~\ref{fig:bar-nii} we show an \NII{} image of the bright bar
region in the Orion nebula that shows evidence of just such a
structure.%
\footnote{This image is based on data obtained with the
  WFPC2 instrument on the \textit{Hubble Space Telescope}, provided by
  C. R. O'Dell.} 
The bar is believed to be a section of the principal ionization front
in Orion that is seen almost edge-on. It can be seen as a diffuse
strip of \NII{} emission ($\mathrm{width} \simeq 10'' \simeq
\unit{6\times 10^{16}}{cm}$) stretching from the top-left to bottom
right of the image. The principal source of ionizing radiation is the
O7V star \thC{}, located off the image to the NW\@. A sharp, bright
edge to the emission can be seen on its SE side along a considerable
fraction of the bar, which may correspond to the electron density
peak.

The geometry of the nebula in the vicinity of the bar is far more
complicated than the plane-parallel geometry assumed in the models,
making a quantitative comparison difficult. The bar probably consists
of at least two overlapping folds in the ionization front and its
appearance is also affected by protruding fingers of neutral gas and
interactions with the HH\,203/204 and HH\,528 jets. However, the
straight region of the bar to the NE of the HH\,203/204 bowshocks show
a relatively simple structure, which we will attempt to compare with
our model predictions. We present in Figure~\ref{fig:bar-profiles}
emission line spatial intensity profiles along a short section of
narrow slit parallel to the bar, with position as indicated in
Figure~\ref{fig:bar-nii}. Comparison of Figure~\ref{fig:bar-profiles}
with the model profiles of Figure~\ref{fig:ZH007c} shows good general
agreement.

The electron density in the bar region has been measured from the
\SIIbothlam{} line ratio to be around
\unit{4000}{\rpcubic{\centi\meter}}
\citep{1995ApJ...438..784W,garcia03}, whereas model ZH007 predicts a
value of $\simeq \unit{8000}{\rpcubic{\centi\meter}}$ for the
\SII{}-derived density. On the other hand, the incident ionizing flux
can be estimated to be about
$\unit{5\times10^{12}}{\rpsquare{\centi\meter}}$, which is half that
of the model ZH007. Thus the model has the same ionization parameter
as the bar and the results can be directly compared by multiplying the
model lengths by a factor of two such that the spatially broad
component to the \NII{} emission is predicted to have a thickness of
$\simeq \unit{4\times 10^{16}}{\centi\meter} \simeq 7''$, in
reasonable agreement with what is observed. The electron density peak
at the ionization front is predicted to have a thickness of $\simeq
\unit{2\times 10^{15}}{\centi\meter} \simeq 0.3''$, which is also very
close to the observed thickness of the narrow ridge in \NII{} (note
that this thickness is fully resolved by the \textit{HST}, which has
an angular resolution of $\simeq 0.1''$ at optical wavelengths).

\subsection{Velocity-Ionization Correlation in the Orion Nebula}
\label{sec:veloc-ioniz-corr}

The mean velocity of different optical emission lines from the core of
the Orion nebula has long been known to correlate with the ionization
potential of the parent ion  \citep{1967ApJ...148..925K,
  1993ApJ...403..678O, 1999AJ....118.2350H, 2001ApJ...556..203O,
  2004Doi-kinematics}. Lines from more highly ionized species such as
\OIIIlam{} are blue-shifted by approximately \Velocity{10} with
respect to the molecular gas of OMC-1, which lies behind the nebula,
with intermediate-ionization species such as \NIIlam{} being found at
intermediate velocities. 

The results of section~\ref{sec:results} show that just such a
correlation can be qualitatively reproduced by our models. However, on
closer inspection, many significant differences are revealed between
the model results and the Orion observations. A much clearer
velocity-ionization correlation exists in model ZL009 (low ionization
parameter) then in ZH007 (high ionization parameter, more pertinent to
the Orion nebula), as can be seen from Table~\ref{tab:ZL009-lines} and
Figures~\ref{fig:ZL009d} and \ref{fig:ZH007d}. This is not surprising
since the emission in the high ionization parameter model is dominated
by ionized equilibrium gas, where velocity changes are rather small
and are driven by variations in the thermal balance (the complex
interplay of radiation hardening and the excitation of different
coolant lines) rather than being directly caused by ionization
changes. Furthermore, both the magnitude of the velocity shifts seen
in the models and the broadening induced by the gas acceleration are
somewhat smaller than is observed in Orion. 

In order to better reproduce the observations, what is required is a
means of continuing the acceleration of the gas inside the body of
the nebula, where hydrogen is fully ionized. There are two means by
which this might be achieved: first, by including the continuum
radiation force upon the ionized gas/dust mixture, or, second, by
considering a transonic strong-D ionization front in a non-plane,
divergent geometry. 

\begin{figure}
  \includegraphics[width=\linewidth]{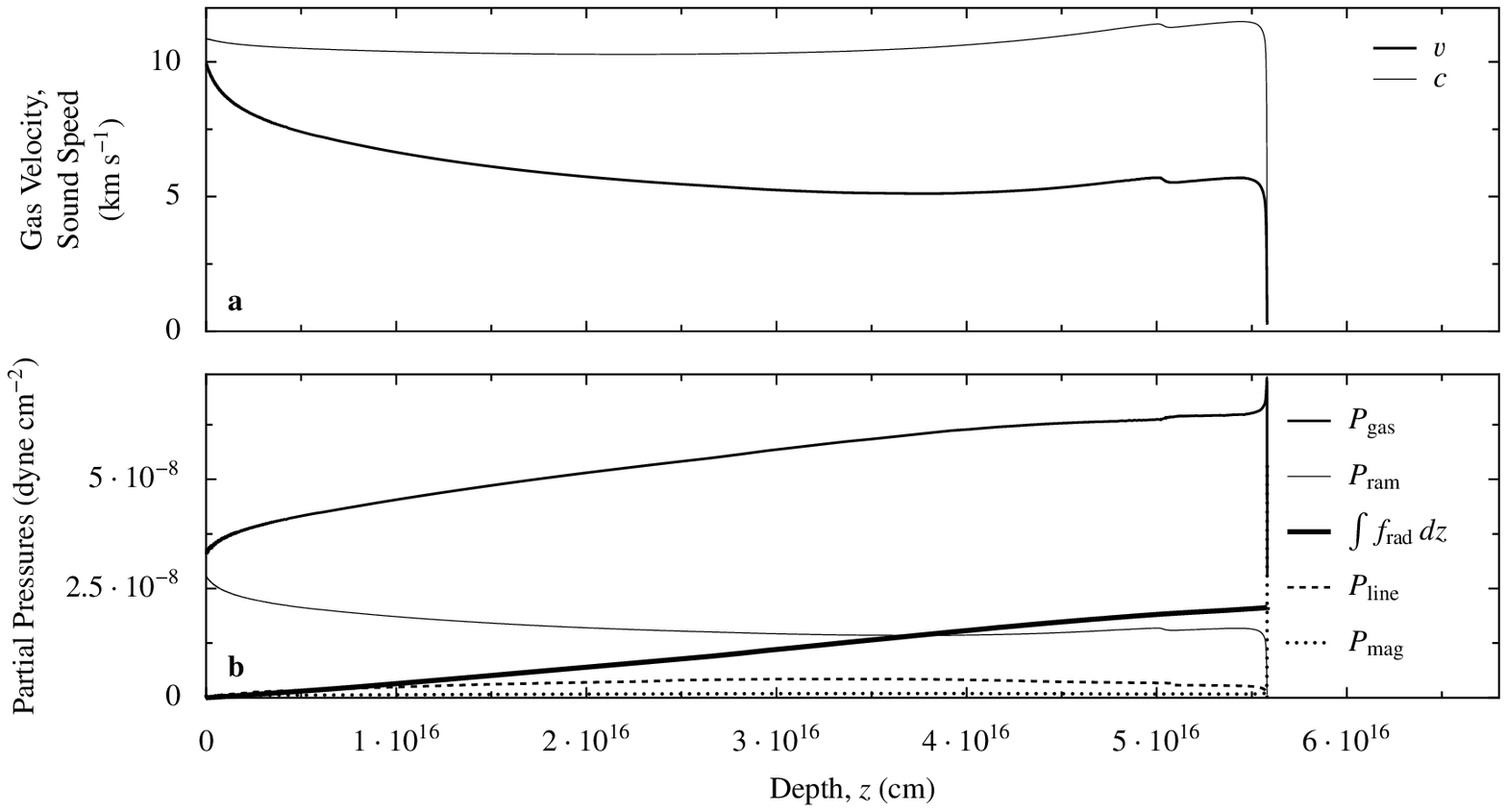}
  \caption{Structure of model ZHR012, which includes the continuum
    radiation force. (\textit{a}) Velocity and isothermal sound speed.
  (\textit{b}) Partial contributions to the total pressure: thermal
  gas pressure (medium weight line), ram pressure (thin line),
  integrated radiative force (thick line), resonance line radiation
  pressure (dashed line), and magnetic pressure (dotted line). }
  \label{fig:ZHR012}
\end{figure}

The results of a model that includes the continuum radiation force are
shown in Figure~\ref{fig:ZHR012}. The thick line in the lower panel
shows the integral along the radiation propagation direction of the
radiative force per unit volume, $f\Sub{rad}$, that acts on the
material in the flow.\footnote{The dust-gas drift velocity is always
  much slower than the flow velocity in these models, so it is valid
  to suppose that the gas and dust dynamics are perfectly coupled.}
In response to this radiative forcing, the total pressure increases
with depth and, since the flow is subsonic, the gas pressure and
density increase in the same direction. By mass conservation, this
leads to an acceleration of the gas towards the illuminated face, as
can be seen in Figure~\ref{fig:ZHR012}\textit{a} (compare
Figure~\ref{fig:ZH007a}\textit{a}, where this process is absent).
 
\begin{figure}
  \includegraphics[width=\linewidth]{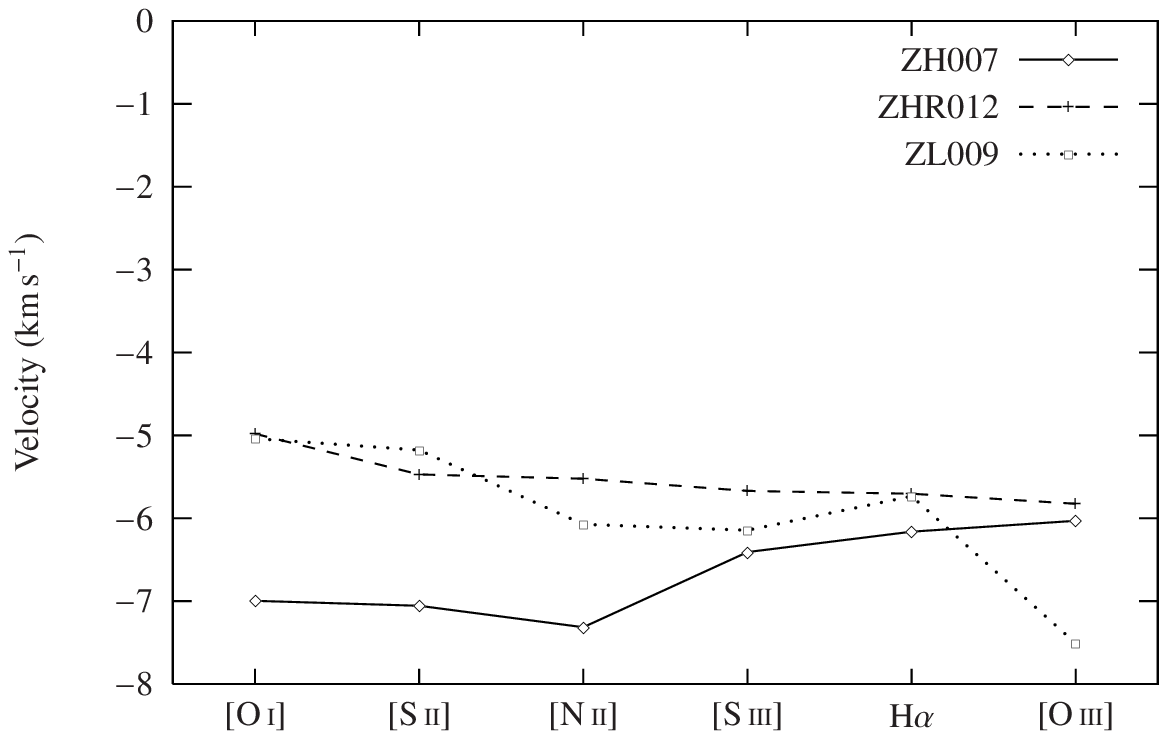}
  \caption{Mean velocities of common optical emission lines from plane
    parallel advective models in face-on orientation. The ionization
    potential of the parent ion increases from left to right.}
  \label{fig:mean-vel}
\end{figure}

However, this acceleration of the fully-ionized gas does not have a
very large effect on the mean velocity of the emission lines, as can
be seen from Figure~\ref{fig:mean-vel}. This is because the
higher-velocity gas represents only a small fraction of the total
emission, even for the \OIII{} line. Although the model ZHR012 does
show a clear relation between velocity and ionization (unlike ZH007),
the gradient is very small, being less than \Velocity{1} between \OI{}
and \OIII{}, compared with an observed difference of 5 to
\Velocity{10} in the Orion nebula. Also, any deviations from a
strictly face-on orientation of the observer would tend to reduce the
observed gradients. 

In order to reproduce the observed kinematics of the Orion nebula, one
needs a strong acceleration of the gas in the region between the
hydrogen and helium ionization fronts, since this is the region where
the ionization of heavy elements such as oxygen and sulfur is changing
most swiftly. We have shown that it is not possible to achieve this
with a plane-parallel model. In such a geometry, gas acceleration
requires either a gradient in the sound speed or the application of a
body force, neither of which are present with the required magnitude
in the relevant region. Strong changes in the sound speed only occur
at greater depths, in the hydrogen ionization front, while the
effective gravity associated with continuum radiation pressure acts
mainly at shallower depths, where the heavy-element ionization is not
changing. 

On the other hand, as we will show in a following paper, the requisite
acceleration is a natural consequence of models in which the flow is
divergent (either spherical or cylindrical) rather than
plane-parallel. Although such models are obviously relevant to such
objects as proplyds \citep{1999AJ....118.2350H} and photoevaporating
globules \citep{1990ApJ...354..529B}, it is not so apparent that they
should apply to the large-scale emission from the Orion nebula, which
has been traditionally visualized as a bowl-like cavity on the near
side of the molecular cloud OMC-1. However, three-dimensional
reconstruction of the shape of the ionization front
\citep{1995ApJ...438..784W} indicates that the radius of curvature of
the front is larger than its distance from the ionizing star. In such
a case, the divergence of the radiation field can lead to a (weaker)
divergence of the flow \citep{2003RMxAC..15..175H}. Another possibility is that
the mean flow is the superposition of multiple divergent flows from
the many bar-like features that have been found in the nebula
\citep{2000AJ....120..382O}.

\section{Conclusions}
\label{sec:conclusions}

In this paper, we have developed a method for including the effects of
steady material flows in the plasma physics code Cloudy
\citep{2000RMxAC...9..153F}, which was previously capable of modeling
only static configurations. We have presented a detailed description
of the numerical algorithms and an example application to the
restricted problem of plane-parallel ionization-bounded \hii{} regions
where the flow does not pass through a sonic point (weak fronts). The
most important conclusions from our study are as follows:

\begin{enumerate}
\item\label{conc:local} The local effects of advection are most
  important in those regions of the flow where physical conditions are
  strongly varying over short distances, such as in the hydrogen
  ionization front.
\item\label{conc:global} The global effects of advection on an \hii{}
  region depend on the relative thicknesses of the ionization front
  and the region as a whole, which is a function of the ionization
  parameter. Only for low values of the ionization parameter, such as
  are found in cometary knots of planetary nebulae, do we find a
  significant direct effect of advection on the emission properties
  integrated over the entire the region. For higher ionization
  parameters, more typical of \hii{} regions around O~stars, the
  effects of advection are indirect and more localized.
\item\label{conc:temperature} One such indirect effect is a
  modification of the temperature structure in the ionization front
  due to the overheating of partially ionized gas. However, we find
  the magnitude of this effect to be much less than has previously
  been claimed \citep{Osterbrock-book,1998A&A...331..347R}, probably
  due to our more realistic treatment of heating and cooling
  processes. For weak D-type fronts (subsonic flows), the temperature
  reached in the front does not exceed the equilibrium temperature in
  the fully ionized gas (see Figure~\ref{fig:ZH007b}\textit{b}). Even
  for supersonic R-type fronts, the peak temperature is only about
  20\% higher than the equilibrium ionized value (see
  Figure~\ref{fig:ZH050b}\textit{b}).  The temperature increase causes
  an increase in the peak emissivity of the \OIlam{} line, but the
  total emission of this line tends to be \emph{less} than in an
  equivalent static model because advection acts to sharpen the
  ionization front and hence decreases the width of the zone where
  \OI{} is emitted.
\item\label{conc:eden} Another indirect effect of advection in high
  ionization parameter regions is the production of a sharp spike in
  the electron density, which occurs at the ionization front whenever
  the peak Mach number in the flow exceeds about 0.5. This spike does
  not occur in static models and its existence can be shown
  analytically to be a consequence of the exchange between thermal
  pressure and ram pressure as the gas is accelerated through the
  front (Appendix~\ref{sec:analytic-model}, Figure~\ref{fig:weakd}).
  As such, it can serve as a useful diagnostic for the presence of
  advection in ionization fronts, best observed in the \NIIlam{} line,
  for which advective models predict a two-component structure
  (Figure~\ref{fig:ZH007c}\textit{b}): a broad peak of emission from
  the equilibrium H$^+$-He$^0$ zone plus a narrower peak from the
  $n\Sub{e}$ spike at the ionization front. Just such a structure is
  seen in \textit{HST} images of the Orion bar
  (Section~\ref{sec:electr-dens-struct}), suggesting that advection is
  important in that region.
\item\label{conc:kinematics} Finally, the advective models provide a
  mapping between the ionization state of the gas and its velocity and
  can hence be used to predict kinematic profiles of different
  emission lines, which can be compared with spectroscopic
  observations. We make such a comparison in the case of the Orion
  nebula but find that the plane-parallel models presented in this
  paper are utterly incapable of explaining the observed kinematics
  (Section~\ref{sec:veloc-ioniz-corr}). The observations seem to
  demand a strong acceleration of the gas in the region between the
  hydrogen and helium ionization fronts, whereas the only acceleration
  mechanisms that can act in the plane-parallel models either occur at
  greater depths (gas heating in the hydrogen ionization front) or at
  shallower depths (effective gravity due to continuum radiation
  pressure).
\end{enumerate}

The work presented in this paper forms a basis for further development
of dynamic photoionization models that will be covered in two
following papers. In the first, the inclusion of chemistry and
dissociation/formation processes allows a unified treatment of the
entire flow from cold, molecular gas, through the PDR, and into the
\hii{} region. Previous work by \citet{1998ApJ...495..853S} indicates
that advection can have a significant effect on the position of the
molecular hydrogen dissociation front. In the second, divergent,
transonic flows from strong-D fronts are modeled, which can explain
the kinematic observations discussed in
paragraph~\ref{conc:kinematics} above.

\acknowledgments

We are most grateful to Bob O'Dell for providing the \textit{HST}
WFPC2 mosaics of the Orion nebula in various optical emission lines
and to Torsten Elwert for useful comments on a draft of the
manuscript.  We are grateful for financial support from NASA
Astrophysics Theory Program Grant NAG 5-12020 and DGAPA-UNAM, Mexico,
grant PAPIIT IN115202.  In addition, WJH and SJA are grateful to the
University of Leeds, UK, for hospitality during a year's sabbatical
visit, supported by grants from DGAPA-UNAM.  RJRW acknowledges support
from a PPARC Advanced Fellowship while this work was developed, and
thanks Cardiff University for hosting him during the period of this
grant. RJRW also thanks the Ministry of Defence for support while this
work was completed.




\clearpage

\appendix

\section{Simplified Analytic Model for Weak-D Ionization Fronts}
\label{sec:analytic-model}

In this appendix, we develop a simple analytic model for a
plane-parallel ionization front in order to explore the most important
effects of advection upon the front structure. The principal
simplification involved is the assumption that the gas temperature,
$T$, follows a unique prescribed function of the hydrogen ionization
fraction, $x$. It is a generic property of ionization fronts that the
temperature tends to increase as one passes from the neutral to the
ionized side of the front. In the analytic model, we assume that the
increase is monotonic and has the form:
\begin{equation}
  \label{eq:tlaw}
  T(x) = T\Max -  
  \frac{1-x^{\beta\Sub{T}}}{1+x^{\beta\Sub{T}} }
  (T\Max-T_0), 
\end{equation} where $T\Max$ is the limiting temperature in the
ionized gas ($x \rightarrow 1$), $T_0$ is the limiting temperature on
the far neutral side ($x\rightarrow 0$), and $\beta\Sub{T}$ is a
parameter controlling the sharpness of the transition.  In reality,
for moderate to high ionization parameters, the hardening of the
radiation field as one approaches the ionization front causes the
temperature to have a maximum for $x$ somewhat less than unity (see,
for example, Figure~\ref{fig:ZH007b}\textit{b}). However, for weak-D
fronts, equation~(\ref{eq:tlaw}) is sufficient to
capture the main effects of
advection.\footnote{%
  For D-critical/strong-D fronts, on the other hand, the position of
  the temperature maximum is critical.}%
Also, it turns out that advection itself will modify the $T(x)$ curve
(see Section~\ref{sec:results}) but, again, we ignore this
complication in the analytic model. We further simplify the model by
only considering the ionization of hydrogen and ignoring the effects
of radiation pressure, dust, and magnetic fields.

\begin{figure}
  \includegraphics[width=\linewidth]{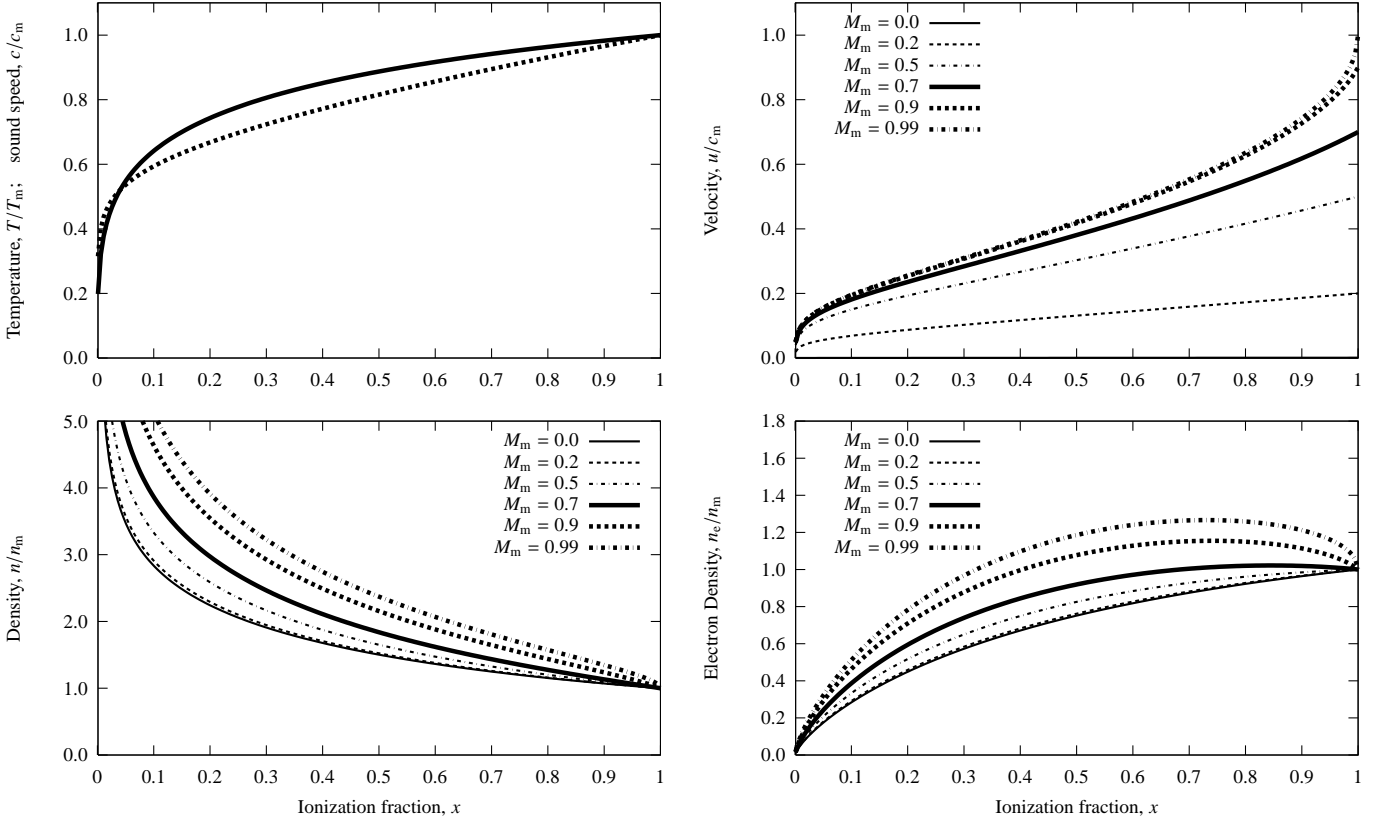}
  \caption{Weak-D solutions with various maximum Mach numbers
    $\Mach{}\Max$}
  \label{fig:weakd}
\end{figure}

With these approximations, the gas pressure is given by 
\begin{equation}
  \label{eq:press-equilib}
  P = n (1 + x ) k T , 
\end{equation}
where $n$ is the hydrogen number density. For a static front, the gas
pressure will be constant, so it is a simple matter to calculate the
variation of gas density, $n$, and electron density, $n\Sub{e}$, with
ionization fraction:
\begin{equation}
  \label{eq:densities-static}
  \frac{n(x)}{n\Sub{m}} = \frac{2}{1+x}
  \left(\frac{T(x)}{T\Sub{m}}\right)^{-1}; 
  \qquad 
  n\Sub{e}(x) = x n(x).
\end{equation}
These are all plotted as solid lines in Figure~\ref{fig:weakd}. In
this and all following examples, we use temperature-law parameters of
$\beta\Sub{T} = 1/3$ and $T_0/T\Sub{m} = 0.02$, which agrees within
10--20\% with the temperature profiles of all the weak-D models in
section~\ref{sec:results}. It can be seen that, although the gas
density declines with increasing $x$, the electron density is a
monotonically increasing function of $x$ (this is always true,
whatever the value of $\beta\Sub{T}$).

In order to calculate the structure of a non-static, dynamic front it
is necessary to consider the conservation of mass and momentum, which
are given, in plane-parallel geometry, by
\begin{equation}
  \label{eq:conservation}
    n m\subH v = \Phi_0  \quad\mathrm{and}\quad P + n m\subH  v^2 = \Pi_0 , 
\end{equation}
where $v$ is the gas velocity and $\Phi_0$ and $\Pi_0$ are the
(constant) mass and momentum fluxes. It is convenient to define the
dimensionless variable $\phi$: 
\begin{equation}
  \label{eq:phi-definition}
  \phi \equiv \frac{ \Pi_0 } { 2 c \Phi_0 } ,  
\end{equation}
which, by equations~(\ref{eq:conservation}), is related to the
isothermal Mach number ($\Mach{} \equiv
v/c$) via 
\begin{equation}
  \label{eq:phi-of-M}
  \phi = \frac{1}{2} \left( \Mach{} + \frac{1}{\Mach{}}
  \right) , 
\end{equation}
with the inverse relation 
\begin{equation}
  \label{eq:M-of-phi}
  \Mach{} = \phi \pm \left( \phi^2 - 1 \right)^{1/2} ,  
\end{equation}
in which the $-$ sign applies to a subsonic flow and the $+$ sign to
supersonic flow. 

Equation~(\ref{eq:phi-of-M}) shows that $\phi \ge 1$ everywhere and
that $\phi = 1$ corresponds to the sonic point: $\Mach{} = 1$. A given
ionization front can be characterized by the parameter $\phi\Max$,
which is the minimum value of $\phi$ anywhere in the front. This is
achieved when the isothermal sound speed $c$ attains its
\emph{maximum} value $c\Max$, and, in the case of the weak-D fronts
considered here, corresponds to the maximum value of the
(subsonic) Mach number: $\Mach{}\Max$.\footnote{%
  For R-type fronts, on the other hand, $\Mach{}\Max$ is the minimum
  Mach number. For D-critical and strong-D fronts, $\Mach{}\Max = 1$
  and is no longer a turning point in \Mach{}, not even a point of
  inflexion, although
  $\phi\Max$ is still a maximum in $\phi$.} %
For a temperature profile such as equation~(\ref{eq:tlaw}), $c\Max$
occurs at $x = 1$ and the sound speed varies with $x$ as
\begin{equation}
  \label{eq:c-law}
  c(x) = c\Max \left( \frac{T(x)}{T\Max} \frac{1+x}{2} \right)^{1/2} . 
\end{equation}  

It can be seen that all weak-D fronts with a given $T(x)$ form a
one-parameter family, characterized by their maximum Mach number
$\Mach{}\Max$.  For a given $\Mach{}\Max$, the corresponding
$\phi\Max$ can be calculated from (\ref{eq:phi-of-M}) and then from
(\ref{eq:phi-definition}) we have $\phi(x) = \phi\Max c\Max / c(x)$,
which can be inserted into (\ref{eq:M-of-phi}) to give $\Mach{}(x)$,
from whence we also have $v(x) = \Mach{}(x) c(x)$, $n(x) = n\Max
\Mach{}\Max c\Max / v(x)$, and $\Ne{}(x) = x n(x)$. Hence, the full
structure of the ionization front as a function of $x$ can be found
algebraically. The results are plotted in Figure~\ref{fig:weakd} for
various weak-D fronts between $\Mach{}\Max = 0$ and $0.99$.

For $\Mach{}\Max = 0.2$, the velocities are everywhere very subsonic
and the density structure is hardly any different from the static case
(lower panels). In this regime, the velocity rises approximately as $u
\simeq c^2 \Mach{}\Max / c\Max \propto (1+y) T(y)$.  There is hence an
initial brisk acceleration for $y < 0.05$, driven largely by the
increase in $T$, followed by a slower, almost constant, acceleration,
driven largely by the increase in ion fraction. For higher
$\Mach{}\Max$, the gas density contrast between the ionized and
neutral sides increases and for $\Mach{}\Max > 0.7$ this causes there
to be a maximum in \Ne{} at an intermediate value of $y$. Solutions
with $\Mach\Max > 0.7$ also show a second episode of steep
acceleration as $y \rightarrow 1$.

In order to apply these results, it is necessary to find the mapping
between the ionization fraction, $x$, and physical position within the
ionization front. For this, it is necessary to introduce more
parameters than have so far been considered. The ionized gas is
assumed to have an inner boundary $z = 0$, at which the ionizing flux
is $F_0$, and with $z$ being the distance into the ionized gas,
measured from the illuminated face, in the direction of decreasing
$x$. Globally, the flux of ionizing photons at $z = 0$ must be
balanced by the recombinations per unit area, integrated throughout
the structure, \emph{plus} the (constant) flux of hydrogen nuclei
through the front: 
\begin{equation}
  \label{eq:global-ion-balance}
  F_0 = \frac{\Phi_0}{m\subH} + \int_0^\infty \alpha(x) x^2 n^2\, dz 
\end{equation}
where $\alpha$ is the recombination coefficient, which we approximate
as a power-law in the gas temperature: $\alpha = \alpha\Max
(T/T\Max)^{-1}$.  The global importance of advection on the ionization
balance can be characterized by the relative magnitude of the two
terms on the RHS of this equation, so we define a dimensionless
``global advection parameter'':
\begin{equation}
  \label{eq:def-lambda}
  \lamad \equiv \frac{\Phi_0}{F_0 m\subH - \Phi_0} .
\end{equation}
We also define a characteristic ``Strömgren distance'' as
\begin{equation}
  \label{eq:z0}
  z_0 \equiv \frac{1+\lamad}{\alpha\Max n\Max^2} \int_0^\infty \alpha x^2 n^2\, dz 
  ,
\end{equation}
in terms of which, (\ref{eq:global-ion-balance}) becomes
\begin{equation}
  \label{eq:global-ion-balance-2}
  F_0 = \alpha\Max n\Max^2 z_0 . 
\end{equation}

At each point, $z$, within the structure we also have the \emph{local}
ionization equation:
\begin{equation}
  \label{eq:local-ion-balance}
  \alpha x^2 n^2 - F (1-x) n \sigbar  = \frac{d}{ d z} (xnv) 
  = \frac{\Phi_0}{m\subH} \frac{d x}{ d z},
\end{equation}
where $F$ is the local value of the ionizing flux and $\sigbar$ is
the mean photo-absorption cross-section for ionizing photons,
evaluated by integrating over the ionizing spectrum at that point. The
attenuation of the ionizing radiation is expressed by 
\begin{equation}
  \label{eq:dF-dz}
  \frac{d F}{d z} = -\sigbar n (1-x) F. 
\end{equation}

Equations~(\ref{eq:local-ion-balance}) and (\ref{eq:dF-dz}) can be
reexpressed in dimensionless form as 
\begin{equation}
  \label{eq:dx-dtaubar}
  \frac{d x}{d \bar{\tau}} = \frac{1}{\Mach\Max \xiad}
  \left[ \frac{\tilde{\alpha}\tilde{n}x^2}{\tilde{\sigma}(1-x)} 
    - \tau_* e^{-\bar{\tau}} \right] ,
\end{equation}
and
\begin{equation}
  \label{eq:dzilde-dtaubar}
  \frac{d \tilde{z}}{d \bar{\tau}} = \left[ \tau_*
  \tilde{\sigma}\tilde{n}(1-x) \right]^{-1} , 
\end{equation}
with dimensionless variables:
\begin{displaymath}
  \tilde{z} \equiv \frac{z}{z_0}, \qquad
  \tilde{n} \equiv \frac{n}{n\Max}, \qquad
  \tilde{\alpha} \equiv \frac{\alpha}{\alpha\Max}, \qquad
  \tilde{\sigma} \equiv \frac{\sigbar}{\sigbar_0}, \qquad
  \bar{\tau} \equiv  - \ln \frac{F}{F_0} .
\end{displaymath}
Equations~(\ref{eq:dx-dtaubar}) and (\ref{eq:dzilde-dtaubar}) also make
use of two dimensionless  parameters:
\begin{equation}
  \label{eq:tau-star}
  \tau_* \equiv n\Max \sigbar_0 z_0 , 
\end{equation}
and
\begin{equation}
  \label{eq:xiad}
  \xiad \equiv \frac{c\Max\sigbar_0}{\alpha\Max},  
\end{equation}
in terms of which the global advection parameter, $\lamad$, can be
expressed as 
\begin{equation}
  \label{eq:lamad-xiad-taustar-mach}
  \lamad = \frac{ \xiad \Mach\Max } { \tau_* - \xiad \Mach\Max } , 
\end{equation} and whose significance is explored more fully in
section~\ref{sec:advect-ioniz-fronts}. These two differential
equations can be integrated numerically to find the full solution for
the ionization front structure in physical space. Note that for the
static case ($\Mach\Max = 0$), (\ref{eq:dx-dtaubar}) is undefined and
we instead have simply that the term in square brackets is zero.

We now present sample results from solving
equations~(\ref{eq:dx-dtaubar}) and (\ref{eq:dzilde-dtaubar}) using
parameters appropriate to an \hii{} region illuminated by an O~star.
We approximate the reduction in photoabsorption cross-section due to
the hardening of the ionizing radiation field as $\tilde{\sigma}
\simeq \left[ 1 + 0.2 \left( \bar{\tau} + \bar{\tau}^2
  \right)\right]^{-1}$. This is a good fit to the exact result from
assuming the ionizing spectrum of a \unit{40\,000}{\kelvin} blackbody
and $\sigma(\nu) \propto \nu^{-3}$, from which we also obtain
$\sigbar_0 = 0.505\, \sigma(\nu_0) \simeq
\unit{3\times10^{-18}}{\square{\centi\meter}}$. We also assume
$\alpha\Max = \unit{2.6\times10^{-13} }{ \rpcubic{\centi\meter}
  \usk\reciprocal \second}$, $c\Max =
\unit{10}{\kilo\meter\usk\reciprocal\second}$, which give $\xiad =
11.5$. For the parameter $\tau_*$, we adopt the values $30$ and
$3000$, corresponding to a low and a high ionization parameter,
respectively, with the second being more representative of typical
\hii{} regions. The results are shown in Figure~\ref{fig:wdz}. In each
case, curves of the electron density and gas velocity are shown for
models with (right to left) $\Mach\Max = 0.0$, $0.2$, $0.5$, $0.7$,
$0.9$, $0.99$. For $\tau_* = 30$ the direct effects of advection are
considerable, since $\lamad = 0.64 \Mach\Max$, so that for the higher
Mach numbers a substantial fraction of the incident ionizing flux is
consumed by ionization of new atoms, which pushes the ionization front
to the left.

It can also be seen that the advection decreases the width of the
ionization front. In the static front, the width is determined by the
mean free path of ionizing photons in the partially ionized gas, which
gives $\delta z \sim (n\sigma)^{-1}$, whereas in the advective fronts
the ionization fraction at a given value of the ionizing flux is
smaller, leading to a sharper front (4 times narrower in the
near-critical case). Also, the electron density in the advective
models has a peak at the ionization front, which is not seen in the
static models. For $\tau_* = 30$, we have $\lamad = 0.0038 \Mach\Max$,
so the direct effects of advection on the global properties of the
model should be very small.  Nonetheless, the ionization front
position varies by about 5\% between the static model and the
$\Mach\Max = 0.99$ model, which is due to the effect of the electron
density peak.

In order to investigate the effects of advection on the emission-line
properties of the nebula, we consider a generic 
recombination line with emissivity 
\begin{equation}
  \label{eq:emissivity-recombination}
  \eta\Rec(x) = A\Rec \Ne n_{j+1} \, T^{\beta_\eta},
\end{equation}
where $n_{j+1}$ is the number density of the recombining ion, and
a generic collisionally excited line with emissivity
\begin{equation}
  \label{eq:emissivity-collisional}
  \eta\Col(x)  =  \frac{A\Col \Ne n_j}{\displaystyle 1 + B\Col \Ne T^{-1/2}} \, 
  \frac{e^{-E/kT}}{ \sqrt{T} } , 
\end{equation}
where $E$ is the excitation energy of the upper level and $B\Col$ is
the collisional de-excitation coefficient.  The ion density can be
assumed to be $n_j \propto \Ne = xn$ for singly-ionized ions and $n_j
\propto (1-x)n$ for neutral atoms.

\begin{figure}\centering
  \includegraphics[width=\linewidth]{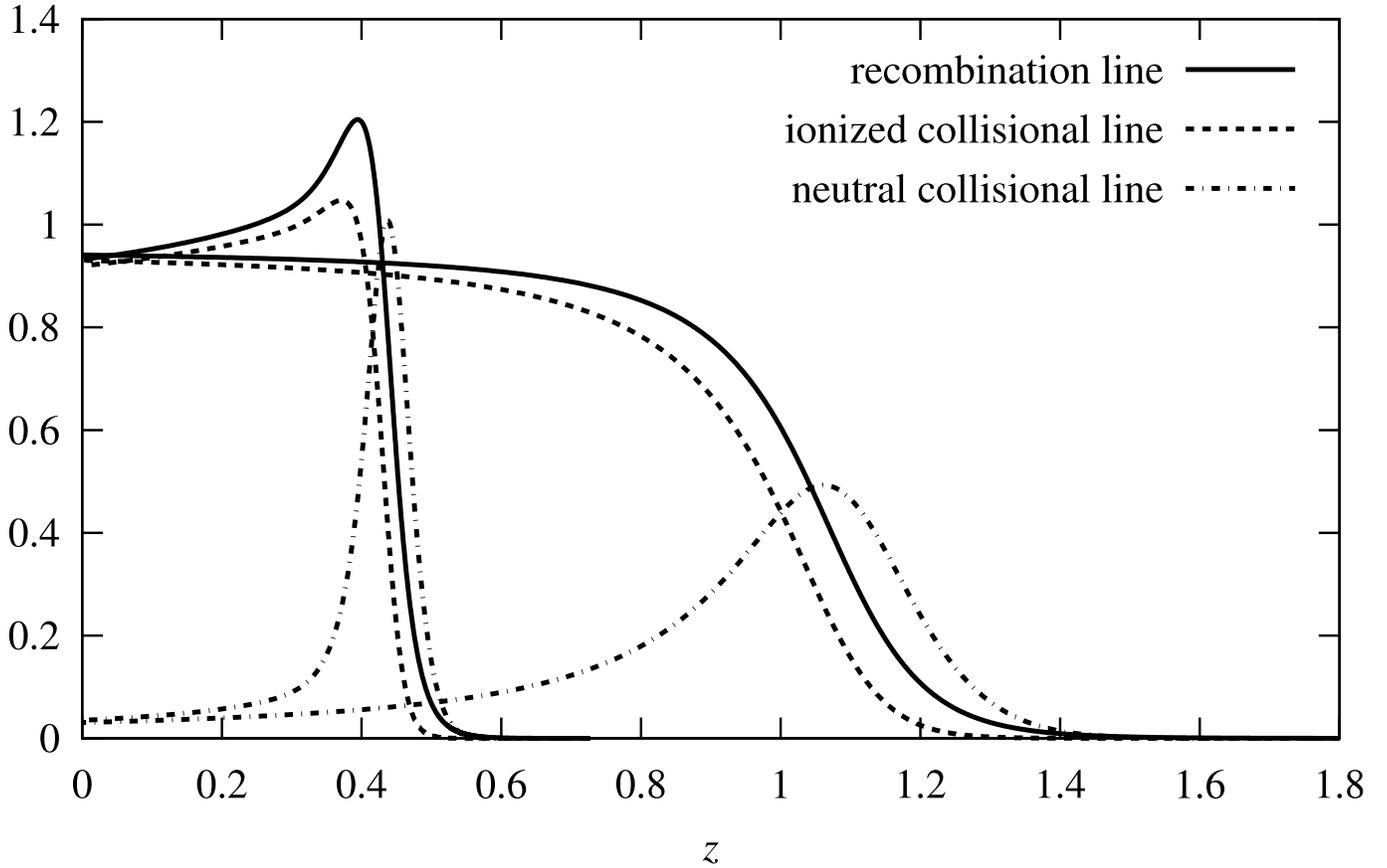}
  \caption{Line emissivity structure of static and dynamic ionization fronts ($\tau_* = 30$),
    in which different line types corresponding to a generic
    recombination line (solid), optical collisional line of an ionized
    species (dashed), and of a neutral species (dotted). The set that
    peaks to the right is from a static model ($\Mach\Max = 0.0$)
    while the set that peaks to the left is from a nearly D-critical
    model ($\Mach\Max = 0.99$). }
  \label{fig:wdz-emissivity}
\end{figure}

The results are shown in Figure~\ref{fig:wdz-emissivity}, which
compares the line emissivities as a function of radius for the $\tau_*
= 30$ model at two values of $\Mach\Max$: 0.0 and 0.99. The total
emission from all the lines is significantly reduced in the nearly
D-critical model with respect to the static model due to the smaller
depth of the ionized zone. On the other hand, the relative intensities
integrated over the entire structure change by less than 10\%. The
peak of the neutral collisional line is much sharpened in the
advective model due to the narrowing of the ionization front. In
addition, the singly-ionized collisional line and the recombination
line both show peaks in their emissivity at the ionization front, which are due
to the electron density peak there (see Figure~\ref{fig:wdz}). 

The line emissivity can be combined with the velocity structure of the
front to create synthetic emission line profiles. We assume that the
front is observed face-on from the ionized side, so that the lines are
all blue-shifted with respect to the neutral gas, which is assumed to
be stationary. We calculate the emergent intensity profile, $I(u)$, of
an emission line ignoring any optical depths effects: 
\begin{equation}
  \label{eq:line-profile}
  I(u) = 
\int_0^\infty \eta(z) \, \exp\left[ -\frac{ \left( v(z) - u
      \right)^2 }{ 2\Delta^2(z) } \right] \, dz , 
\end{equation}
where $u$ is the observed velocity and $\Delta$ is the thermal Doppler
width, calculated at each point in the structure assuming typical
atomic weights of 16 for the collisional neutral line, 14 for the
collisional ionized line, and 1 for the recombination line.

\begin{figure}\centering
  \makebox[\linewidth][l]{(\textit{a})}\\
  \includegraphics[width=\linewidth]{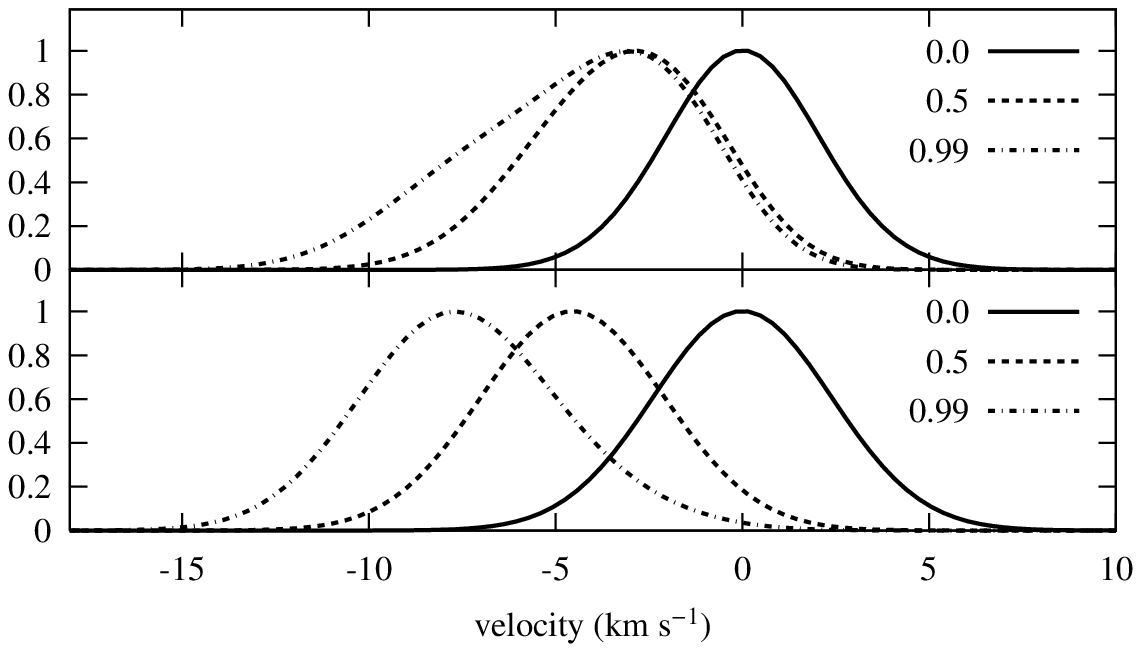}\\
  \makebox[\linewidth][l]{(\textit{b})}\\
  \includegraphics[width=\linewidth]{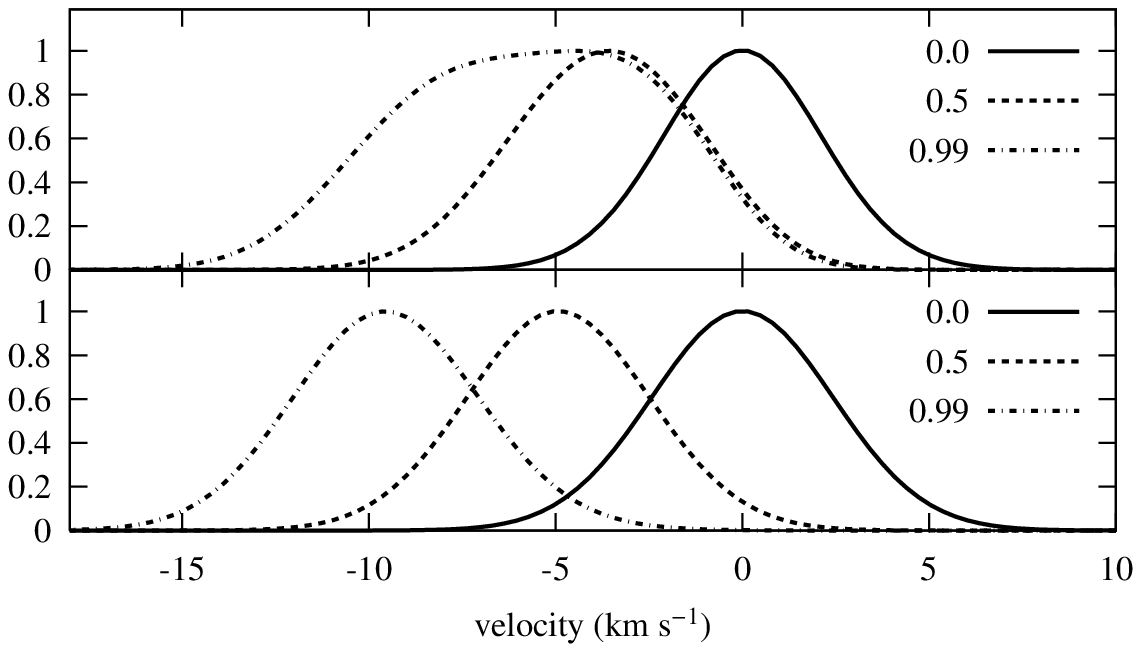}
  \caption{Predicted line profiles from analytic ionization front
  model for (\textit{a}) $\tau_* = 30$ and (\textit{b}) $\tau_* =
  3000$. Top panels shows a generic optical collisional line of a neutral species
  while bottom panels show the same for a singly-ionized
  species. Different line types correspond to $\Mach\Max = 0.0$ (solid
  line), $0.5$ (dashed line, and $0.99$ (dotted line).} 
  \label{fig:wdz-lines}
\end{figure}

The resultant profiles are shown in Figure~\ref{fig:wdz-lines} for the
collisional lines.\footnote{
  The recombination line has a very similar emissivity profile to the
  singly ionized collisional line and the low atomic weight only
  serves to smear out the details of the line profile.}  The
increasing blue-shift of both lines with increasing maximum Mach
number is evident.  For the singly-ionized line, there is hardly any
dependence of linewidth on the advection strength because most of the
emission comes from the almost fully ionized gas, which shows only
small velocity gradients. For the neutral line, on the other hand, the
nearly critical model shows a much broader, double-humped line. The
redder component of the line is due to the emissivity peak around
$x=0.5$ and is little changed between $\Mach\Max = 0.5$ and $\Mach\Max
= 0.99$, whereas the bluer component in the $\Mach\Max = 0.99$ model
comes from the nearly fully ionized gas and is hence relatively
stronger in the model with the higher ionization parameter ($\tau_* =
3000$).

\begin{figure}\centering
  \makebox[\linewidth][l]{(\textit{a})}\\
  \includegraphics[width=0.8\linewidth]{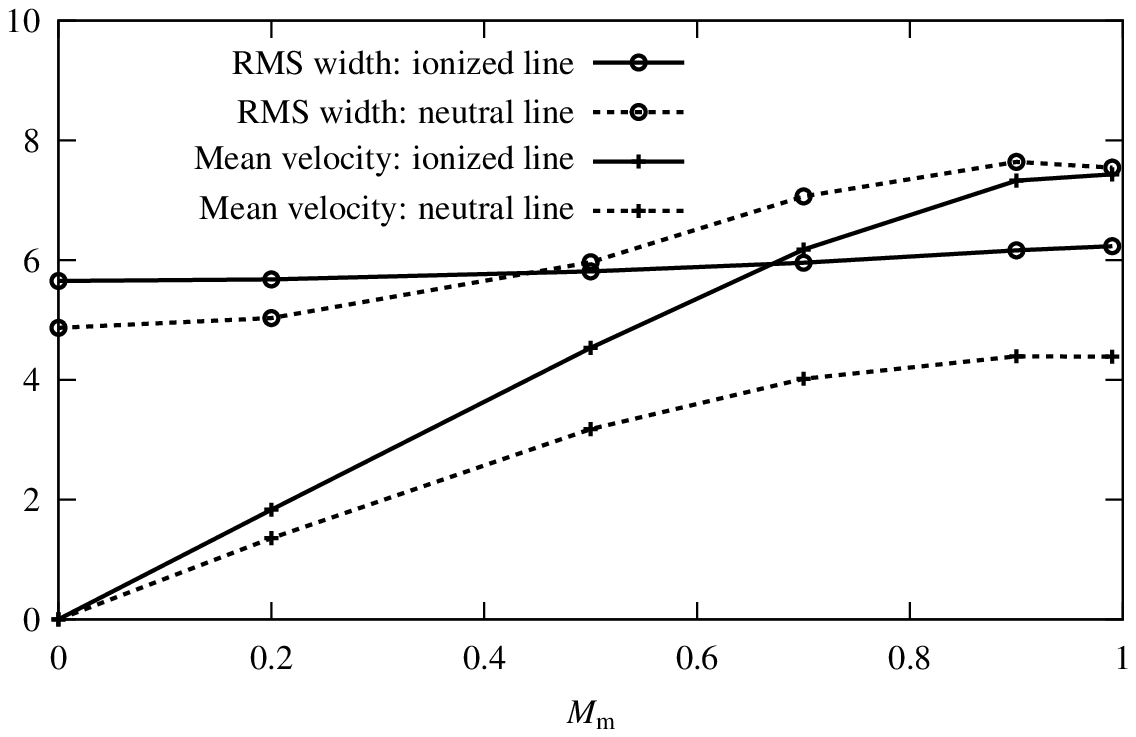}\\
  \makebox[\linewidth][l]{(\textit{b})}\\
  \includegraphics[width=0.8\linewidth]{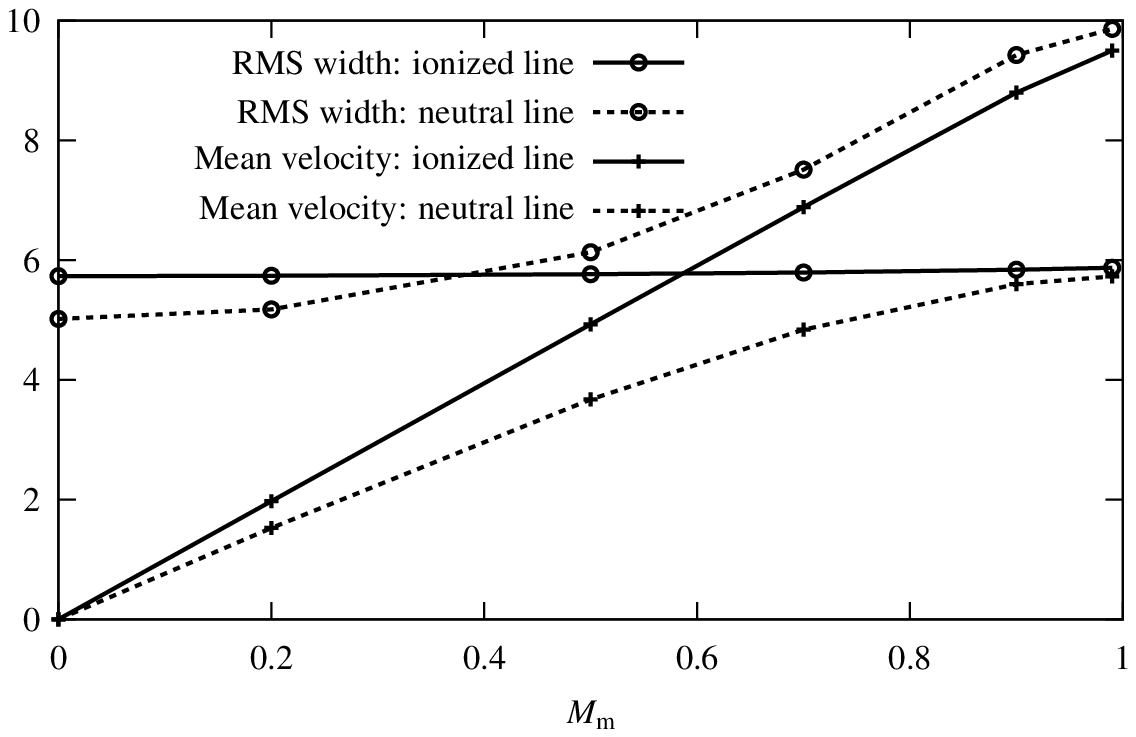}
  \caption{Predicted mean velocities and RMS widths (in
    \kilo\meter\usk\reciprocal{\second}) of collisionally excited
    lines from the analytic ionization front model as a function of
    $\Mach\Max$ for (\textit{a}) $\tau_* = 30$ and (\textit{b})
    $\tau_* = 3000$.}
  \label{fig:wdz-linestats}
\end{figure}

The behavior of the mean velocity, $\bar{v}$, and RMS\footnote{%
  For a Gaussian line profile the full-width-half-maximum, $\Delta v$,
  is related to the RMS width by $\Delta v = 2.306 \sigma$.}  velocity
width, $\sigma$, of the collisional lines as a function of $\Mach\Max$
is shown in Figure~\ref{fig:wdz-linestats}.

\mathcode`\|="8000              
{\catcode`\|=\active \gdef|#1{_{\rm #1}}}
\def\rA{{\rm A}}
\def\rB{{\rm B}}
\def\rC{{\rm C}}
\def\rH{{\rm H}}
\section{Treatment of ionization ladders}
\label{sec:impl-deta}

The rate of change of the fractional abundance of a particular
ionization state is given by
\begin{equation}
{\partial n_i\over \partial t} = G_i+\sum_{j\ne i} R_{j\to i} n_j 
        - n_i\left(S_i+\sum_{j\ne i} R_{i\to j}\right), \label{e:ladder}
\end{equation}
where the $R_{i\to j}$ are the rates for ionization (where $j$ is a
higher state than $i$) and recombination (where $j$ is a lower state
than $i$).  $G_i$ and $S_i$ cater for processes not included within
the ionization ladder, and are respectively the source of ions from
such processes and the sink rate into them.

In previous versions, Cloudy treated the ionization ladders in
isolation, with $G_i = S_i = 0$, and $R_{ij} \ne 0$ only for processes
coupling neighboring ionization states,
\begin{equation}
R_{i\to j} = \left\{
\begin{array}{ll}
{\cal R}_i & j=i+1\\
{\cal I}_j & i=j+1\\
0          &\mbox{otherwise}
\end{array}
\right.
\end{equation}
In this case, equation~(\ref{e:ladder}) in equilibrium ($d/dt\to0$)
gives the equations
\begin{eqnarray}
0 &=& n_2{\cal R}_1 - n_1{\cal I}_1 \\
0 &=& n_3{\cal R}_2 + n_1{\cal I}_1 - n_2({\cal I}_2+{\cal R}_1)\\
0 &=& n_4{\cal R}_3 + n_2{\cal I}_2 - n_3({\cal I}_3+{\cal R}_2)\\
0 &=& n_{N-1}{\cal I}_{N-1} - n_N{\cal R}_{N-1}
\end{eqnarray}
for the abundances $n_i$ of an $N$-state ionization ladder.  These
equations yield a simple expression for the relative abundances of
neighboring ionization states,
\begin{equation}
n_{i+1}/n_i = {\cal I}_i/{\cal R}_i.\label{e:simple}
\end{equation}
The overall abundance of each ionization level can be found using this
relation together with a sum rule for the conserved total abundance of
the species.

This analysis does not apply if we require a time-dependent solution,
or there are more complex interactions between levels (such as the
Auger effect), or external sources and sinks of ions (resulting, for
example, from molecular processes).  

If we assume a general form for all these additional terms, we are
left with a computationally expensive $N\times N$ matrix problem to
solve.  However, the largest coefficients in the matrix derived from
equation~(\ref{e:ladder}) will either be the ionization and
recombination rates, for which we know that a simple solution is
possible, or possibly the time-dependent terms (as, for instance, in
non-equilibrium cooling behind a shock).  This suggests that we should
be able to find a solution to the ladder equations efficiently using
iterative techniques.

We rewrite equation~(\ref{e:ladder}) as
\begin{equation}
\sum_j A_{ij} n_j = \sum_j (\hat{A}_{ij}+\tilde{A}_{ij}) n_j
        = b_j.
\end{equation}
We separate the matrix $A_{ij}$ into two parts, a tridiagonal
component $\hat{A}_{ij}$ and the remainder $\tilde{A}_{ij}$.  The
components of $\tilde{A}_{ij}$ will in general be far smaller than
those of $\hat{A}_{ij}$, so we can use the iterative scheme
\begin{equation}
{\bf n}^{n+1} = 
\hat{A}^{-1}\left({\bf b}-\tilde{A}.{\bf n}^n\right) =
{\bf n}^n+\hat{A}^{-1}\left({\bf b}-A.{\bf n}^n\right)
\label{e:iter}
\end{equation}
to converge to the solution to the ionization state.  In particular,
in the nonlinear system we are treating, the coefficients in $A$ and
${\bf b}$ will themselves be functions of the ionization state of the
gas, and so it suffices to take a single step of the iterative
scheme~(\ref{e:iter}) before these values are updated.

There is one problem with this treatment.  In the limit of small
advection, $\hat{A}^{-1}$ becomes singular.  In this limit, the
solution we require is the null eigenvector of $A$, and as in the
previous treatment we can set its magnitude using an additional
normalization constraint.  However, the rounding error in the
summation of ionization and recombination terms on the diagonal of $A$
can lead to the numerical solution of equation~(\ref{e:iter}) having
negative abundances for states which are substantially less abundant
than their neighbors.  The ease with which the
solution~(\ref{e:simple}) is found suggests that this is not
unavoidable.  By re-writing the standard tridiagonal solver in
\citet{NumericalRecipes}
to treat matrices in the particular form of $\hat{A}$ (and providing
the ionization, recombination and diagonal sink vectors to this
revised algorithm without summation), a near-cancellation is avoided.
The resulting scheme gives solutions which are manifestly positive,
given the physical limits on the signs of the various vector elements.


\section{Magnetic Field}
\label{sec:magnetic-field}
Magnetic field effects can now be included in Cloudy simulations by
specifying the magnetic field strength and geometry at the illuminated
face. Both an ordered field and a `tangled' field may be specified,
although currently the ordered field is restricted to plane-parallel
slab models with advection. The tangled field may be used in any
geometry and with advective or static models.

The tangled field is assumed to provide an isotropic magnetic
pressure. In addition to the field strength at the illuminated face,
$B\Sub{tangled,0}$, the effective magnetic adiabatic index,
$\gamma\Sub{mag}$, must also be specified. This determines the
response of the field to compression of the gas:
\begin{equation}
  \label{eq:mag-tangled-compression}
  B\Sub{tangled} = B\Sub{tangled,0} \left(\frac{\rho}{\rho_0}\right)^{\frac{\gamma\Sub{mag}}{2}}. 
\end{equation}
A value $\gamma\Sub{mag} = 0$ implies a constant magnetic field
strength throughout the model, whereas $\gamma\Sub{mag} = 4/3$ (the
default) corresponds to conservation of magnetic flux and is what
would be expected in the absence of dynamo action or magnetic
reconnection. 

For the ordered field one must specify a component $B_z$ that is
parallel to the integration direction through the slab and a component
$B_t$ that is transverse to the integration direction.  The parallel
component $B_z$ is constant throughout the slab, while the transverse
component is a function of the varying gas velocity, $v$:
\begin{equation}
  \label{eq:mag-transverse-field}
  B_t = B_{t,0} \,\frac{ v_0 - {v\Sub{A}^*} } { v  - {v\Sub{A}^*} } ,
\end{equation}
where $v\Sub{A}^*$ is a characteristic speed
\citep{2001MNRAS.325..293W}:
\begin{equation}
  \label{eq:mag-alfven}
  v\Sub{A}^* = \frac{B_z^2}{4\pi\rho_0 v_0} . 
\end{equation}

Magnetic pressure is included in the gas equation of state, having the
form
\begin{equation}
  \label{eq:mag-pressure}
  P\Sub{mag} = \frac{B\Sub{tangled}^2}{8\pi}
  + \frac{B_t^2-B_z^2}{8\pi} \quad
  \mathrm{dyne\usk\rpsquare{\centi\meter}} . 
\end{equation}
The magnetic contribution to the enthalpy density is given by
\begin{equation}
  \label{eq:mag-enthalpy-density}
  w\Sub{mag} = \frac{\gamma\Sub{mag}}{\gamma\Sub{mag}-1} \frac{B\Sub{tangled}^2}{8\pi}
  + \frac{B_t^2+B_z^2}{4\pi} \quad
  \mathrm{dyne\usk\rpsquare{\centi\meter}} . 
\end{equation}

\end{document}